\documentclass[preprint2]{aastex} 
\usepackage{graphicx, subfigure, pdflscape, float, amsmath, natbib, epstopdf, grffile}
\usepackage[english]{babel}

\slugcomment{To appear in the \textit{Astronomical Journal}}
\shorttitle{The Solar Neighborhood XXXIV}
\shortauthors{Lurie et al.}

\begin{document}


\title{The Solar Neighborhood. XXXIV. \\ A Search for Planets Orbiting
  Nearby M Dwarfs using Astrometry}

\author{John C. Lurie\altaffilmark{1}}
\affil{Department of Astronomy, University of Washington, Seattle, WA 98195}
\email{lurie@uw.edu}

\author{Todd J. Henry\altaffilmark{1}, Philip A. Ianna\altaffilmark{1} }
\affil{RECONS Institute, Chambersburg, PA 17201}

\author{Wei-Chun Jao\altaffilmark{1}, Samuel N. Quinn, and Jennifer G. Winters\altaffilmark{1}}
\affil{Department of Physics and Astronomy, Georgia State University,
       Atlanta, GA 30302}

\author{David W. Koerner}
\affil{Department of Physics and Astronomy, Northern Arizona University, 
       Flagstaff, AZ 86011}

\author{Adric R. Riedel\altaffilmark{1}}
\affil{Department of Astrophysics, American Museum of Natural History, 
       New York, NY 10034}

\and

\author{John P. Subasavage\altaffilmark{1}}
\affil{United States Naval Observatory, Flagstaff, AZ 86001} 

\altaffiltext{1}{Visiting Astronomer, Cerro Tololo Inter-American
Observatory. CTIO is operated by the Association of Universities for
Research in Astronomy, Inc., under contract to the National Science
Foundation.}


\begin{abstract}

Astrometric measurements are presented for seven nearby stars with
previously detected planets: six M dwarfs (GJ 317, GJ 667C, GJ 581, GJ
849, GJ 876, and GJ 1214) and one K dwarf (BD $-$10
3166). Measurements are also presented for six additional nearby M
dwarfs without known planets, but which are more favorable to
astrometric detections of low mass companions, as well as three binary
systems for which we provide astrometric orbit solutions. Observations
have baselines of three to thirteen years, and were made as part of
the RECONS long-term astrometry and photometry program at the
CTIO/SMARTS 0.9m telescope. We provide trigonometric parallaxes and
proper motions for all 16 systems, and perform an extensive analysis
of the astrometric residuals to determine the minimum detectable
companion mass for the 12 M dwarfs not having close stellar
secondaries. For the six M dwarfs with known planets, we are not
sensitive to planets, but can rule out the presence of all but the
least massive brown dwarfs at periods of 2 -- 12 years. For the six
more astrometrically favorable M dwarfs, we conclude that none have
brown dwarf companions, and are sensitive to companions with masses as
low as 1 $\mathrm{M_{Jup}}$ for periods longer than two years. In
particular, we conclude that Proxima Centauri has no Jovian companions
at orbital periods of 2 -- 12 years. These results complement
previously published M dwarf planet occurrence rates by providing
astrometrically determined upper mass limits on potential
super-Jupiter companions at orbits of two years and longer. As part of
a continuing survey, these results are consistent with the paucity of
super-Jupiter and brown dwarf companions we find among the over 250
red dwarfs within 25 pc observed longer than five years in our
astrometric program.
\end{abstract}

\keywords{astrometry --- planetary systems --- solar neighborhood --- stars: low mass}


\section{Introduction}
\label{sec:intro}

During the brief history of extrasolar planet investigations, our
understanding of the relative populations of different types of
planets has been limited by the observational biases of the techniques
employed. With the advent of sophisticated transit searches and
hypersensitive radial velocity measurements, significant progress has
been made discovering various types of planets that orbit stars with
periods up to a few years. Less progress has been made in discovering
planets in longer orbits, and particularly around nearby M dwarfs,
which account for at least 74\% of the stellar population within 10 pc
\citep{hen06}. M dwarfs offer fertile ground for companion searches,
as \citet{dre13} have inferred that a high fraction of M dwarfs host
terrestrial planets at short orbital periods. Less is known about the
populations of Jupiter-mass planets and brown dwarfs around M dwarfs,
particularly at orbital periods longer than a few years.

To understand how M dwarf planetary systems form and evolve, we must
probe the full regime of companion masses and orbital periods. Transit
techniques are geometrically biased towards companions with small
orbits, while radial velocity techniques are biased towards massive
companions with short periods that exert large gravitational
accelerations on their host stars. Direct imaging techniques are
limited to young, giant planets at large separations. Astrometric
techniques, which measure the positions of stars on the plane of the
sky, are most sensitive to Jovian-type planets in Jovian-type
orbits. While radial velocity observing programs are now becoming
sensitive to such companions \citep{bon13a,mon13}, the astrometric
results presented here have longer observational baselines, of up to
13 years. Furthermore, astrometry can detect companions with a large
range of inclinations and orientations, and allow for the
determination of orbit inclinations and accurate companion masses.

To date the majority of nearby extrasolar planets around M dwarfs have
been discovered by radial velocity searches, which tend to select the
brightest M dwarfs. As discussed in more detail in
$\S$\ref{sec:analysis}, in ground-based imaging programs the brightest
targets generally have the noisiest astrometric residuals due to the
short exposures required and the lack of comparably bright reference
stars. With the exception of GJ 1214, five M dwarfs in our sample were
found to have planets using radial velocity techniques, and are among
the brightest targets in our astrometric program. An extreme case is
the K dwarf BD $-$10 3166, for which we are not sensitive to
sub-stellar companions, but for which we provide the first accurate
parallax. For comparison, we have included six additional M dwarfs
that are less bright, less massive, and closer, and therefore more
favorable to companion detection via astrometry. To calibrate our
analysis, we have also included three confirmed stellar binaries with
clear photocentric perturbations for which we have characterized the
orbits. These binaries were chosen from the roughly two dozen binaries
in our observing program with clear astrometric perturbations because
we have observed multiple orbital periods, and can most accurately
characterize the orbits.

Astrometric solutions for proper motion and parallax are given for
each of the 16 systems targeted, plus orbital solutions for three
binaries.  A detailed analysis of the astrometric residuals is given
to search for companions to the 12 M dwarf systems without close
stellar companions. Periodograms of the astrometric residuals have
been generated, along with detection limits based on simulations of 10
million hypothetical companions to each star. These are the first
results of a larger RECONS\footnote{REsearch Consortium on Nearby
  Stars, \textit{www.recons.org}} survey for companions orbiting more
than 250 red dwarfs within 25 pc for which we have at least five years
of coverage. As observations continue, this sample will grow, further
constraining the population of brown dwarf and super-Jupiter
companions in long period orbits around M dwarfs. Finally, to provide
context for these results we provide a comprehensive list of the 17 M
dwarfs within 25 pc having exoplanets as of 1 July 2014, including the
six targeted in this work.

\section{Observations and Reductions}
\label{sec:observe}

\subsection{Astrometry}

The 0.9m telescope at CTIO is equipped with a 2048 $\times$ 2048
Tektronix CCD camera with 0\arcsec.401 pixel$^{-1}$ plate scale
\citep{jao03}.  Only the center quarter of the chip is used for
astrometric and photometric observations, yielding a 6\arcmin.8 square
field of view.  Astrometric exposures are taken through one of four
filters, $V_\mathrm{J}$ (old), $V_\mathrm{J}$ (new), $R_\mathrm{KC}$,
or $I_\mathrm{KC}$\footnote{The central wavelengths for the
  $V_\mathrm{J}$ (old), $V_\mathrm{J}$ (new), $R_\mathrm{KC}$, and
  $I_\mathrm{KC}$ filters are 5438, 5475, 6425, and 8075 \AA,
  respectively.} (hereafter without subscripts, and the $V$ filters
combined).  Depending on the brightnesses of the science targets,
reference stars, and sky conditions, exposure times vary from 20 to
1200 s for targets with 9 $\leq VRI \leq 19$. For optimal centroiding,
exposure times are set so that either science or reference stars have
maximum peak ADU of $\sim$50,000 (digital saturation occurs at 65,537
ADU). Observations are almost always made within $\pm$ 30 minutes of a
science target's transit to minimize the corrections required for
differential color refraction, as described in \citet{jao05}. Three to
five frames are typically taken each night, depending primarily on the
exposure time required. To enable routine calibration of the science
images, bias and dome flat frames are taken nightly.

Instrument setups for most stars have been kept constant during the 13
years of observations. However, we have used two $V$ filters, dubbed
the ``old'' Tek\#2 $V$ filter ($\lambda_{\mathrm{central}}$ = 5438
\AA, $\Delta\lambda$ = 1026 \AA) and ``new'' Tek\#1 V filter
($\lambda_{\mathrm{central}}$ = 5475 \AA, $\Delta\lambda$ = 1000 \AA
), because the ``old'' filter cracked in 2005 February. The ``new''
$V$ filter was used between 2005 and 2009. The ``old'' $V$ filter was
reinstated in 2009  July after confirming that the crack in the corner
did not significantly affect astrometric residuals. As discussed in
\citet{sub09}, a reliable parallax can be obtained using data from
both filters as long as at least 1-2 years of data (depending on
observing frequency) have been taken in each filter.  Reductions
containing both ``old'' and ``new'' $V$ frames can exhibit offsets
of a several milliarcseconds (mas) in residuals on both axes. This has
been mitigated by choosing close-in reference stars, and only using
frames taken near the meridian. In total, 7 of the 16 systems
discussed in this paper were observed astrometrically in the $V$
filter. Further details about the filters and their effects on the
astrometry can be found in \citet{sub09} and \citet{rie10}.

The paths traced on the sky by science stars result from the
combinations of proper motions and parallactic shifts. Details of the
data reduction process used to separate these motions are given in
\citet{jao05} and \citet{hen06} . Briefly, we (1) use SExtractor
\citep{ber96} to measure centroids, (2) define a six-constant plate
model to find plate constants (given in Equation (4) of
\citealt{jao05}), (3) assume that ensembles of reference stars have
zero mean parallax and proper motion, (4) solve least-square equations
for multi-epoch images (given in Equation (5) of \citealt{jao05}), and
(5) convert from relative parallax to absolute parallax by estimating
the distances of the reference stars photometrically.  Our typical
centering precision is 2.1 - 3.5 mas, or 0.5 - 0.9\% of a pixel,
depending on the filter, with $I$ being the best and $V$ being the
worst. To correct the relative parallax to an absolute parallax,
photometric distances are estimated by comparing $VRI$ colors to
$M_{V}$ for single, main-sequence stars in the RECONS 10 pc sample
\citep{hen97,hen06}. A distance is estimated for each reference star,
and the correction to absolute parallax is then computed using the
weighted mean distance of the entire reference field. The uncertainty
on the correction is determined using Equation (6) in \citet{jao05}.

In the case of a binary with a given combination of magnitude and mass
differences\footnote{As described in \citet{vdk67},
  $\alpha=(B-\beta)a$, where $\alpha$ is the photocentric semi-major
  axis of the orbit of the primary, $B$ is the fractional mass
  ($M_{B}/(M_{A}+M_{B})$), $\beta$ is the relative luminosity
  ($1/(1+10^{(0.4)\Delta m})$), and $a$ is the semi-major axis of the
  relative orbit of the two components. The perturbation we detect
  here is $\alpha$.}, we detect its photocenter orbit around its
barycenter, in addition to the motions due to parallax and proper
motion. Hence, the residuals of our typical binary star's parallax
reduction are significantly offset from zero. In order to get a better
parallax result and calculate the photocenter's orbital elements, we
first carry out a standard reduction for proper motion and
parallax. We then fit a photocentric orbit to the residuals, i.e, we
treat these residuals as a binary orbit, using the techniques
described in \citet{hrt89}. Based on the orbital elements we
calculate, this photocentric orbit in components of R.A. and Decl. is
then removed from the centroids of the science star at each
epoch. Finally, we re-calculate proper motion and parallax using these
corrected centroids. The final ``cleaned'' proper motions and
parallaxes are the values given for the binaries in Table \ref{tab1}.
After one iteration, the parallax errors are reduced to those typical
of similar program stars, and the residuals are significantly reduced
and consistent with the mean errors found for our overall program.

\subsection{Photometry}

$VRI$ photometry was obtained at the CTIO 0.9m using the same
instrumental setup used for the astrometry frames. As for astrometry
observations, bias and dome flat frames are taken nightly for basic
image calibration. All science stars were observed at airmass
$< 1.8 $. Exposure times were chosen to reach a signal-to-noise
ratio (S/N) $>$ 100 for science stars in each of the $VRI$
filters. Combinations of fields that provided 10 or more standard
stars from \citet{lan92,lan07} and/or E-regions from \citet{gra82}
were observed several times each night to derive transformation
equations and extinction curves. Further details of photometric data
reductions, the definition of transformation equations, errors, etc.,
can be found in \citet{jao05} and \citet{win11}.

\subsection{Spectroscopy}
Spectroscopic observations used to the provide spectral types in
Table 2 were made at the CTIO 1.5m using the R-C
spectrograph and Loral 1200 $\times$ 800 CCD camera between 2003 and
2006.  Grating \# 32 was used in first order with a tilt of
15.$^{\circ}$1, and observations were made using a 2{\arcsec}
slit. The order-blocking filter OG570 was utilized to provide spectra
covering the range of 6000-9000 {\AA} with a resolution of 8.6
\AA. For calibration, bias frames, dome flats, and sky flats were
taken at the beginning of each night. Further details regarding the
1.5m spectroscopy program and associated data reduction, including
assignment of spectral types, can be found in \citet{hen04}.


\section{Analysis}
\label{sec:analysis}

\begin{figure*}[ht!]
\centering

{\includegraphics[scale=0.225]
{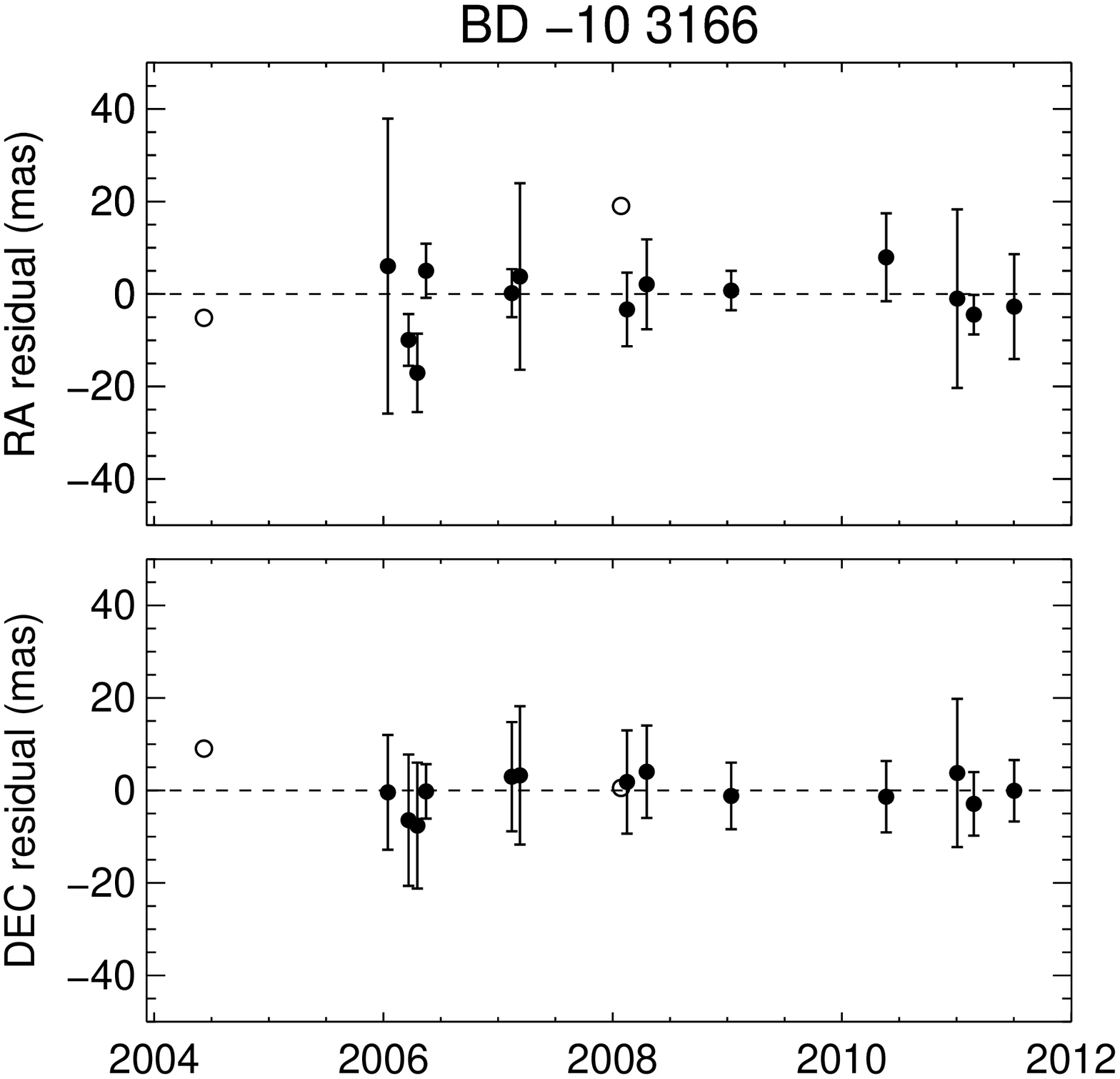}}
\hspace{-0.25in}
\subfigure
{\includegraphics[scale=0.225]
{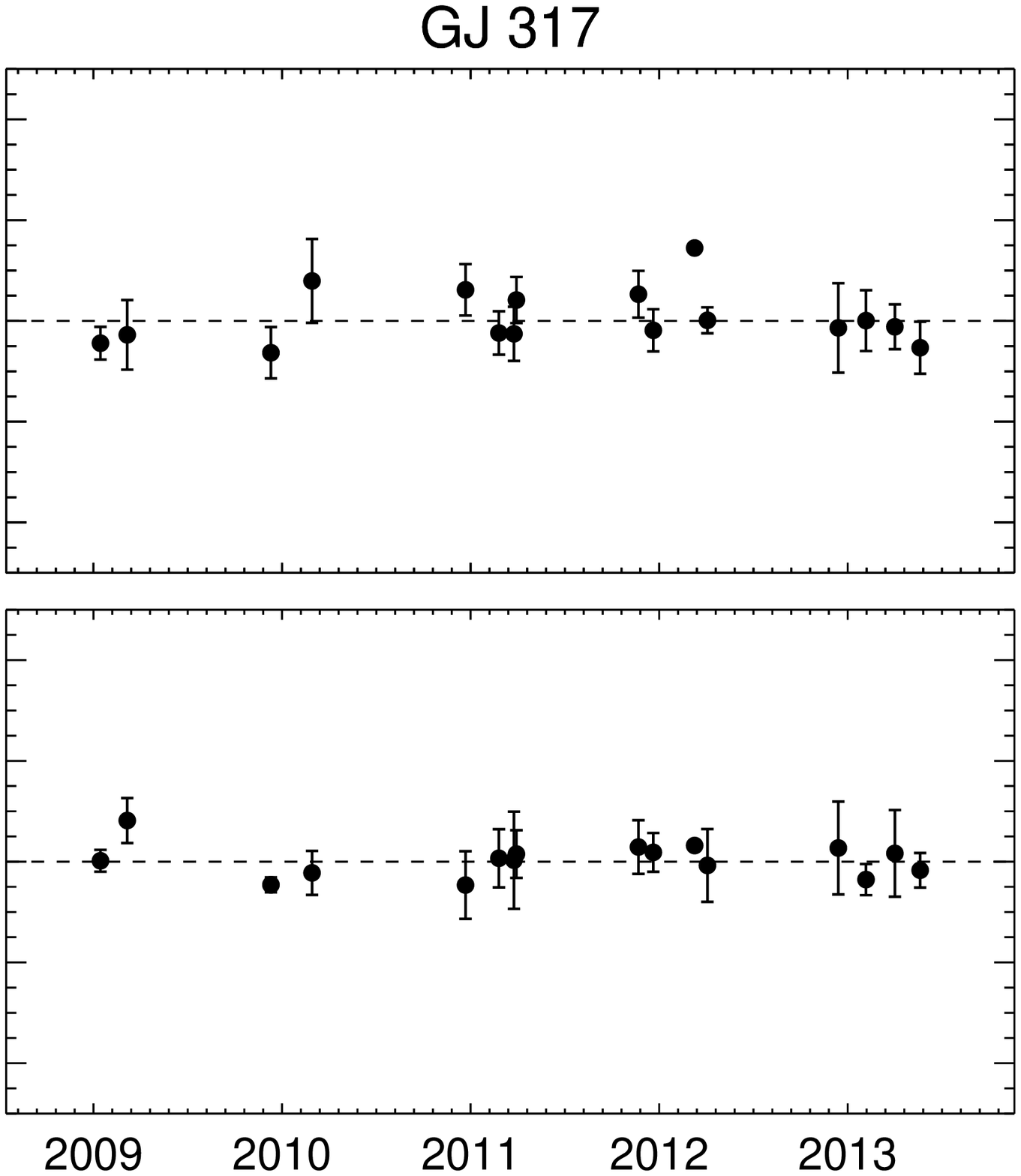}}
\hspace{-0.25in}
\subfigure
{\includegraphics[scale=0.225]
{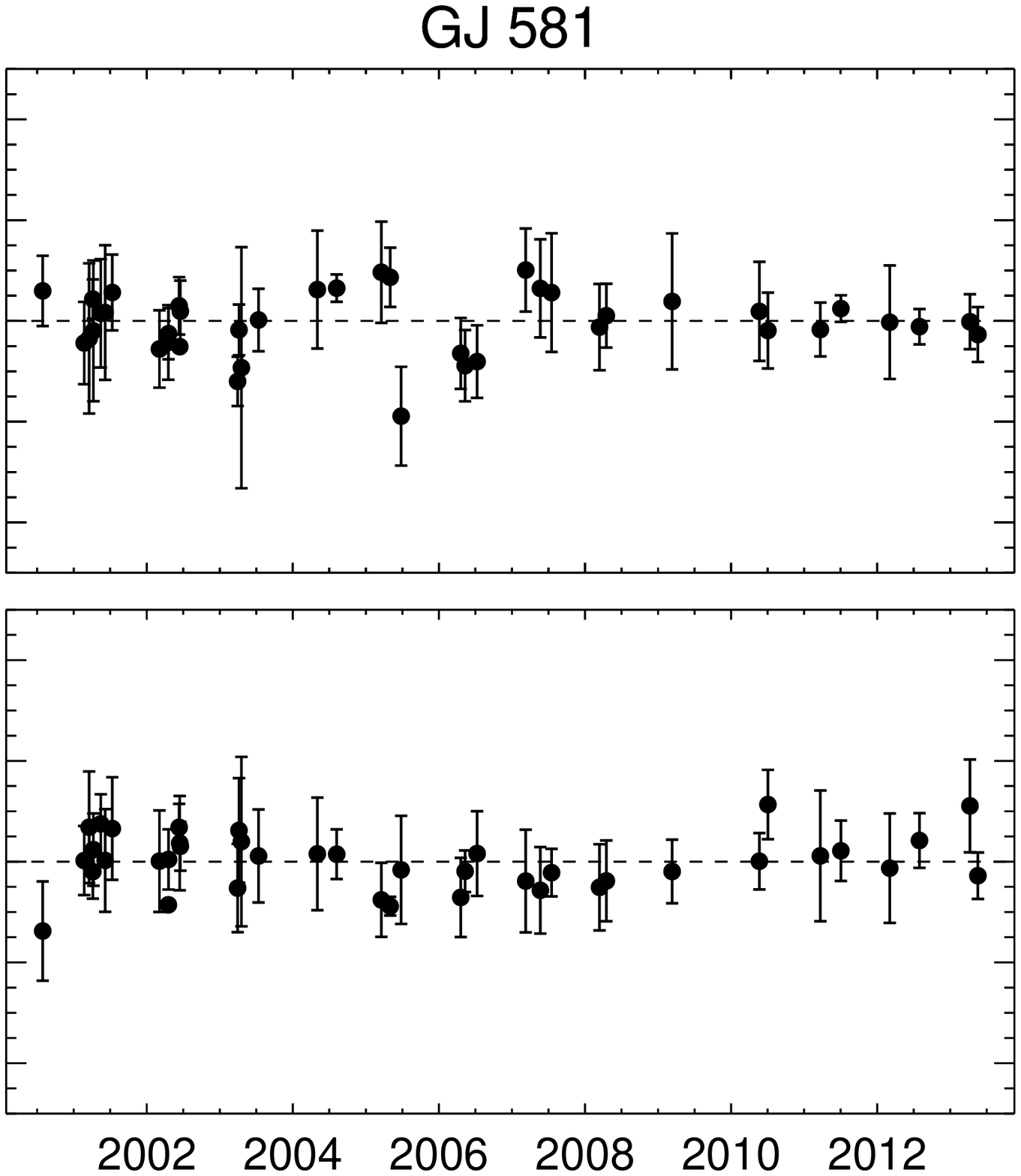}}
\hspace{-0.25in}
\subfigure
{\includegraphics[scale=0.225]
{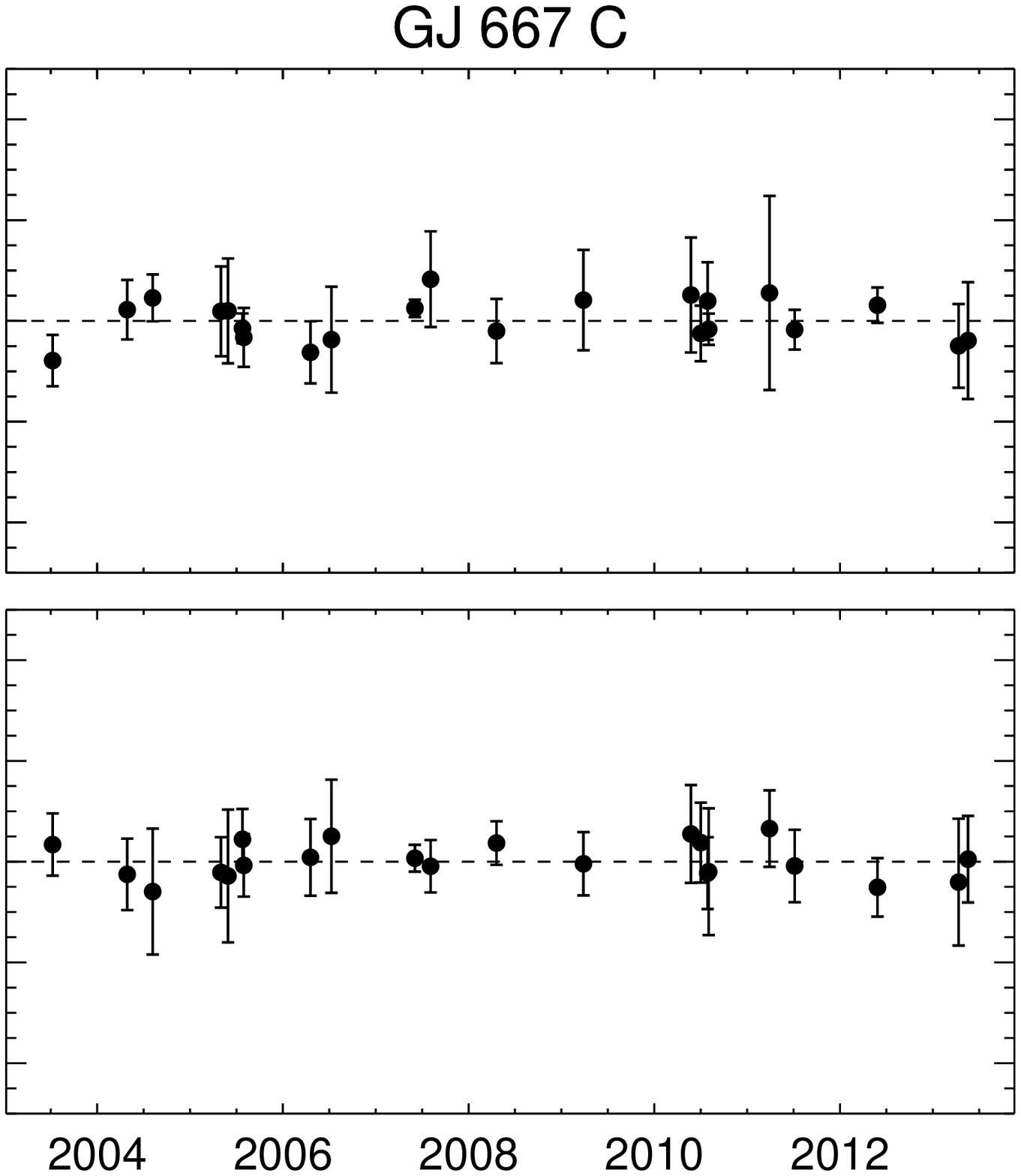}}

\vspace{-.01in}

{\includegraphics[scale=0.225]
{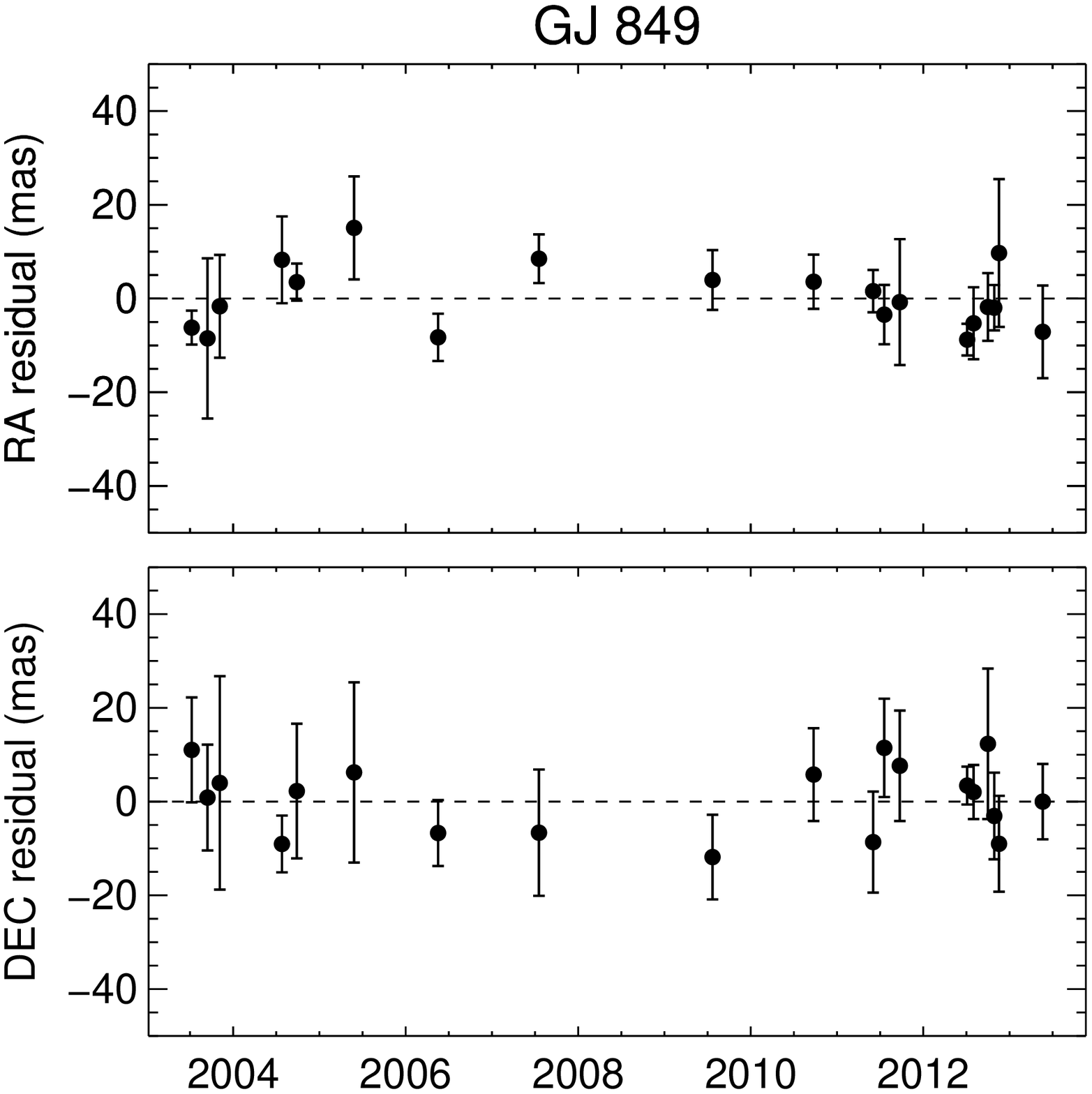}}
\hspace{-0.25in}
\subfigure
{\includegraphics[scale=0.225]
{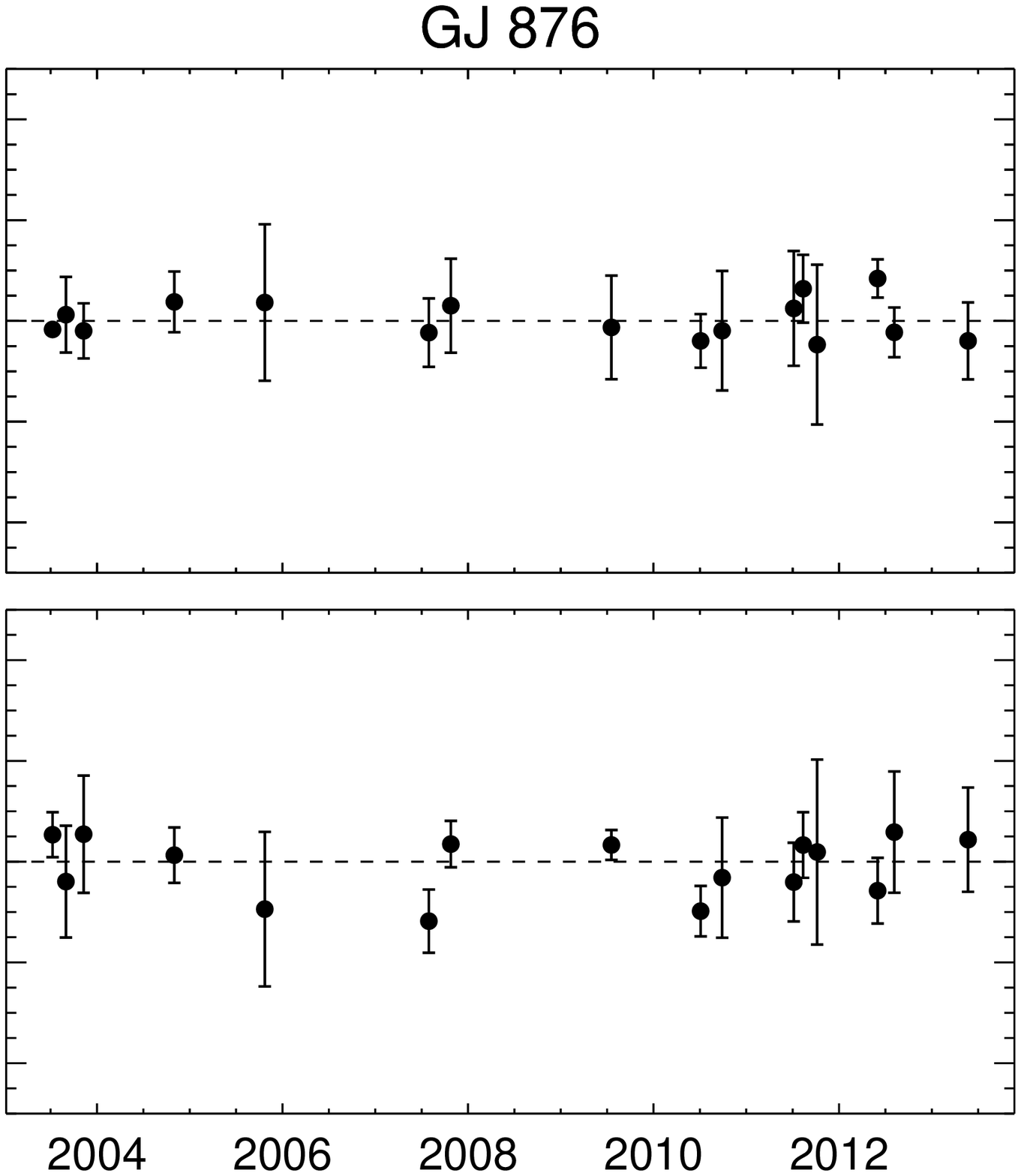}}
\hspace{-0.25in}
\subfigure
{\includegraphics[scale=0.225]
{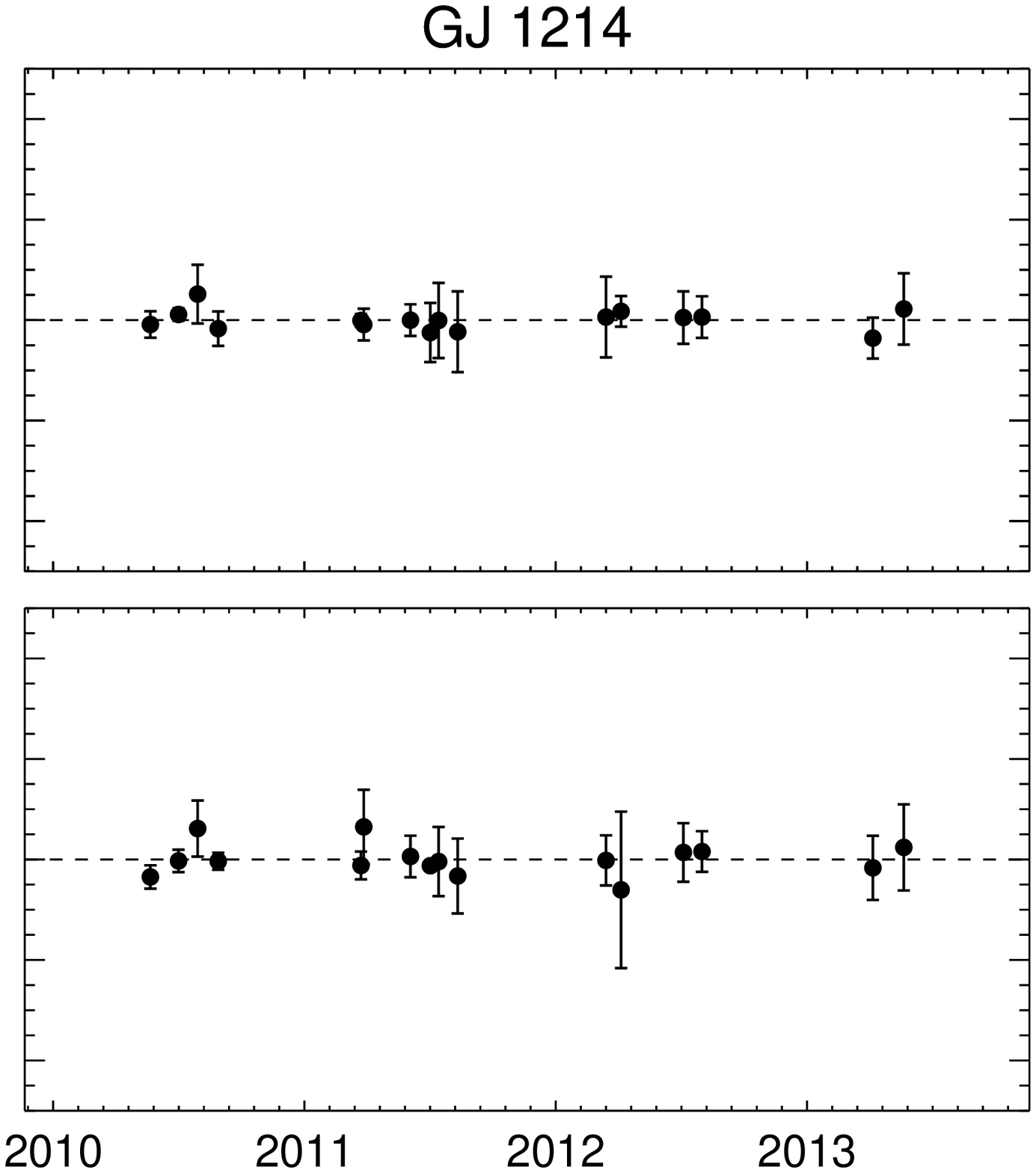}}
\hspace{-0.25in}
\subfigure
{\includegraphics[scale=0.225]
{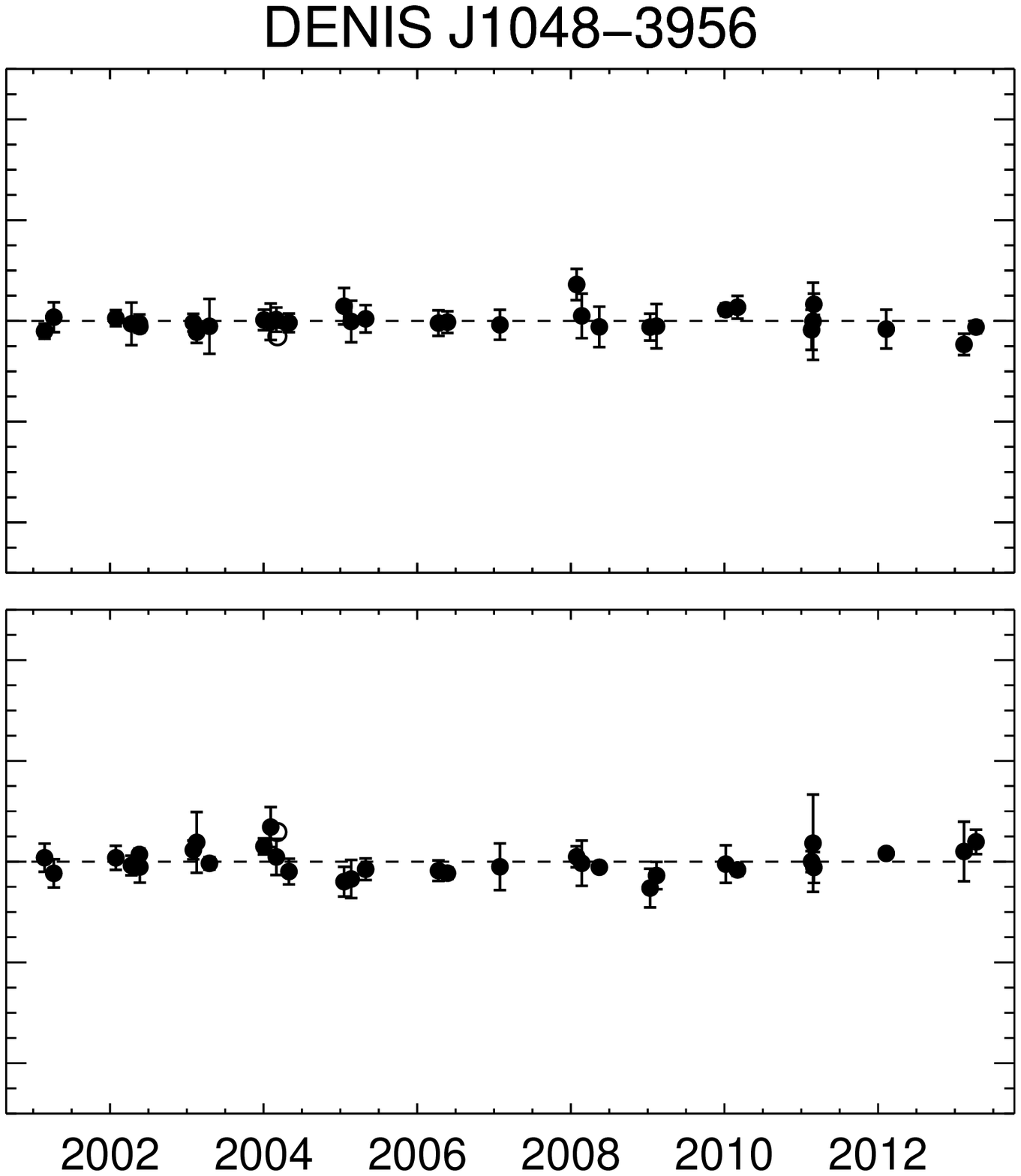}}

\vspace{-.01in}

{\includegraphics[scale=0.225]
{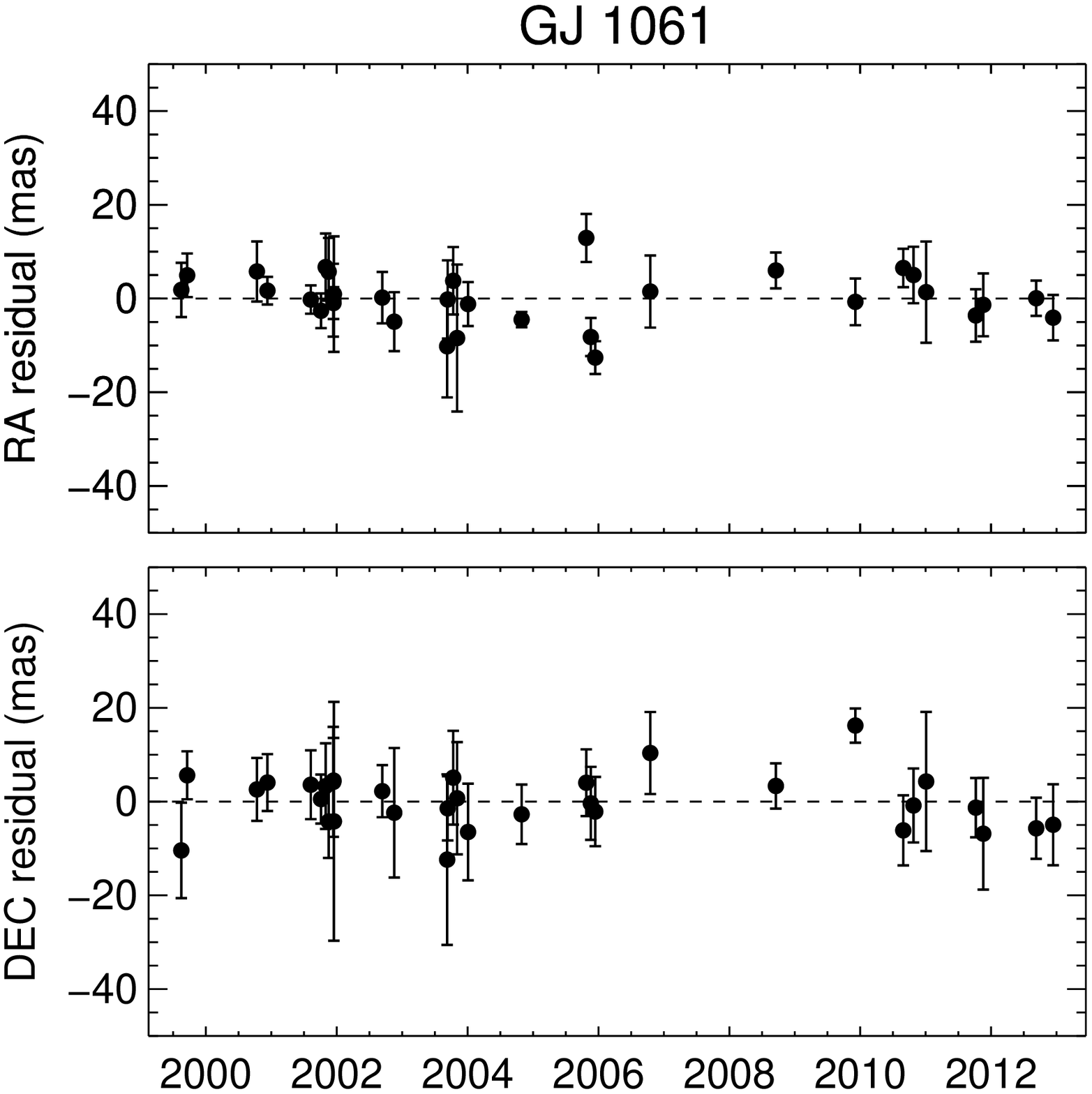}}
\hspace{-0.25in}
\subfigure
{\includegraphics[scale=0.225]
{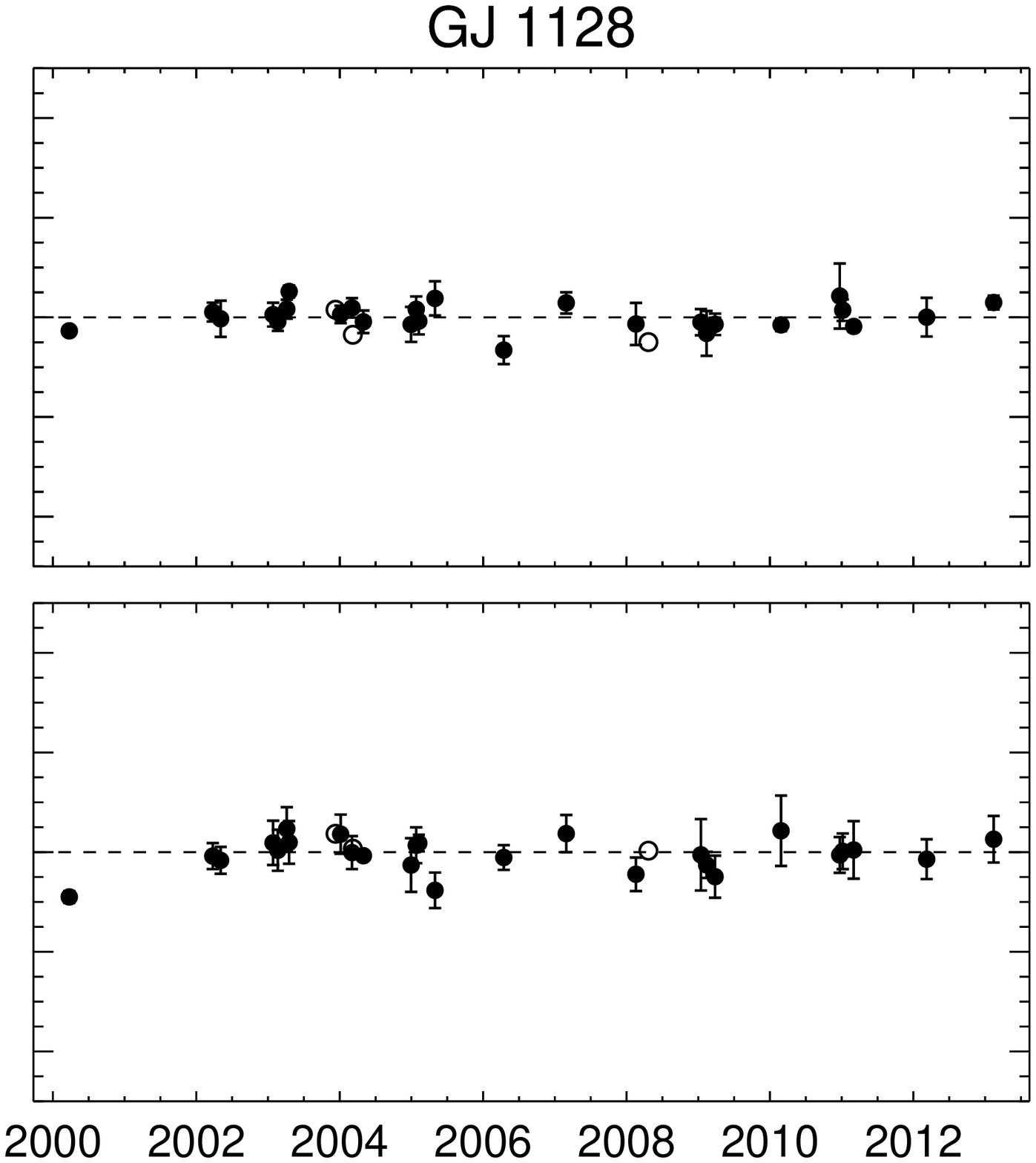}}
\hspace{-0.25in}
\subfigure
{\includegraphics[scale=0.225]
{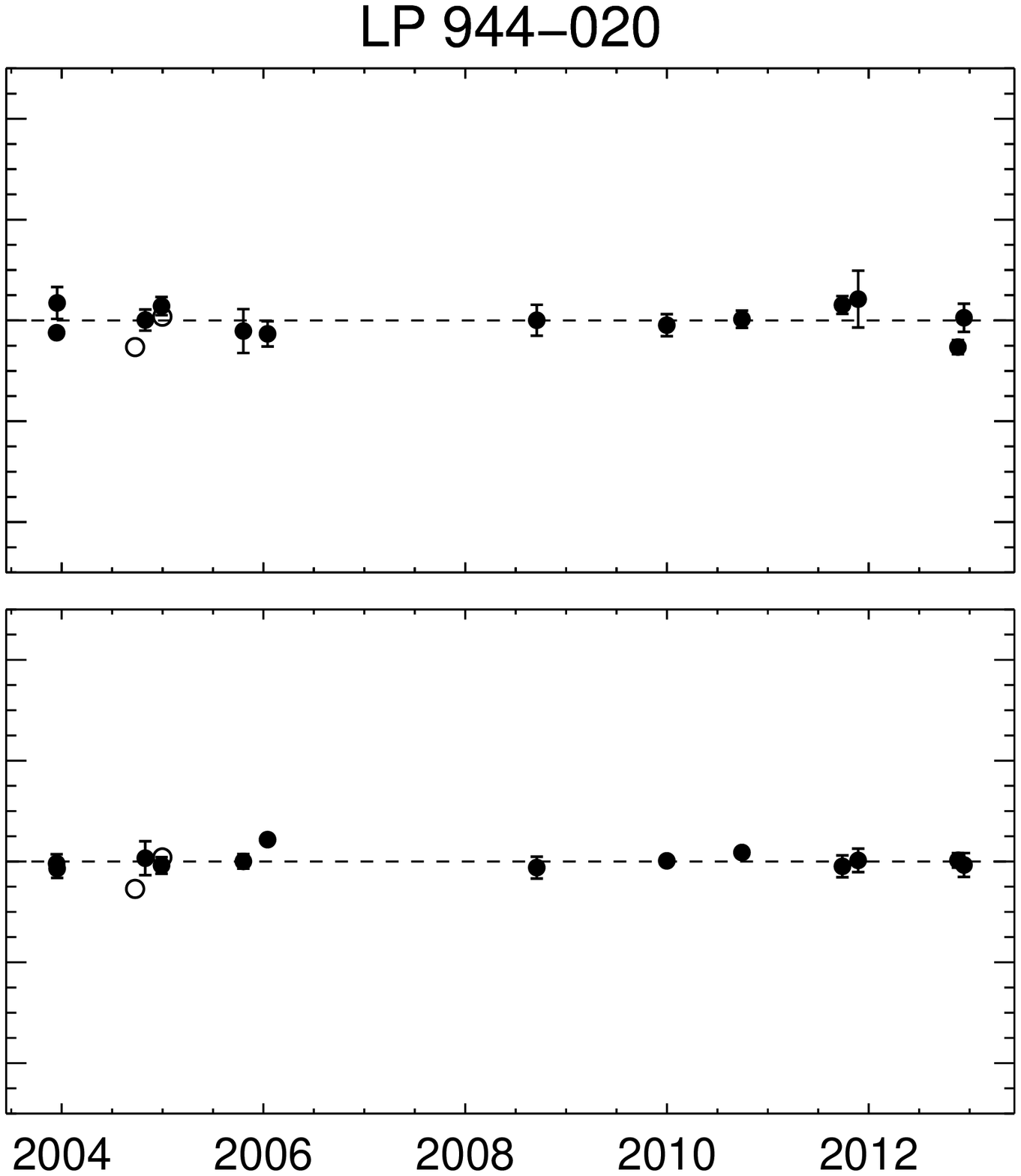}}
\hspace{-0.25in}
\subfigure
{\includegraphics[scale=0.225]
{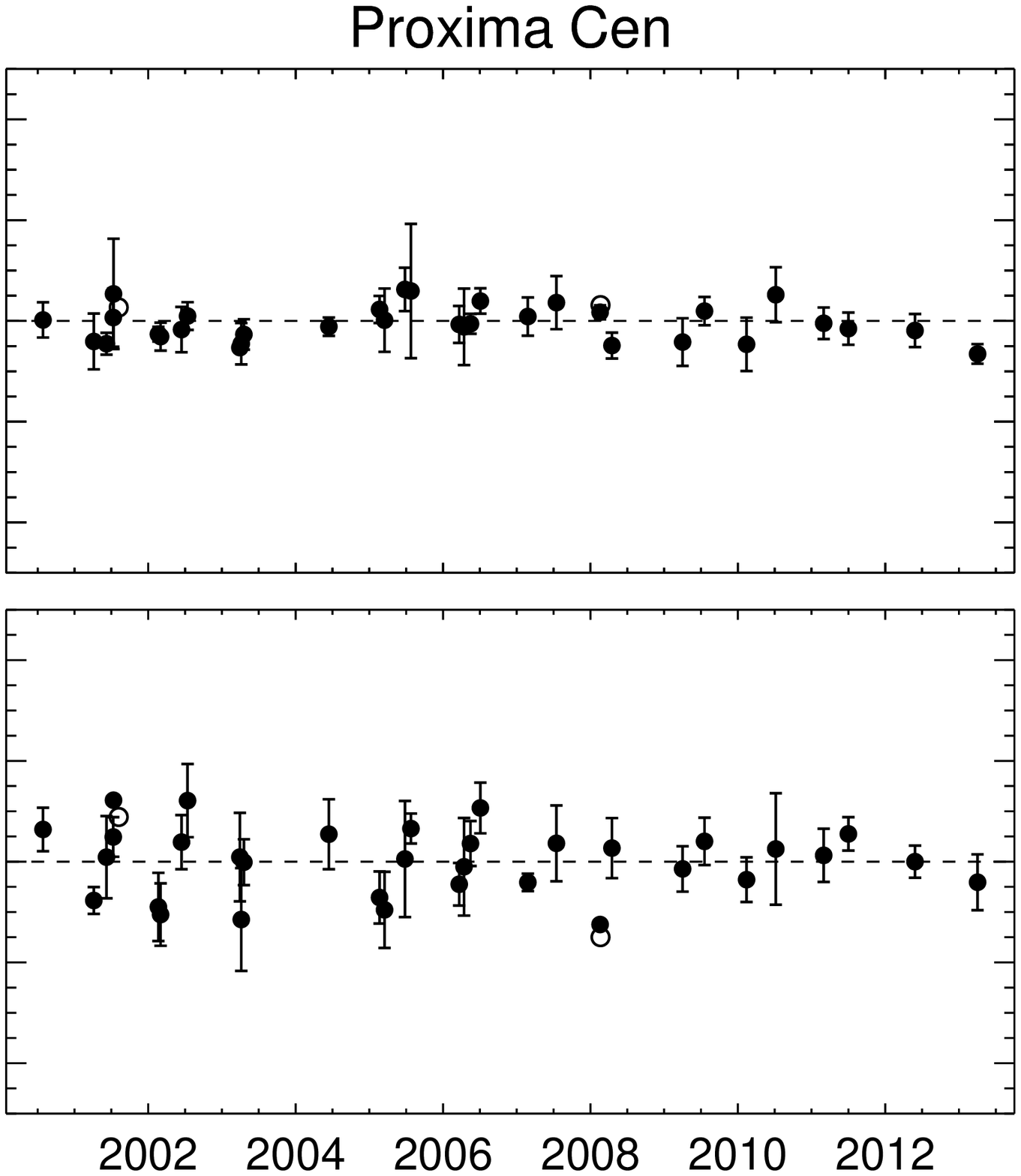}}

\vspace{-.01in}

{\includegraphics[scale=0.225]
{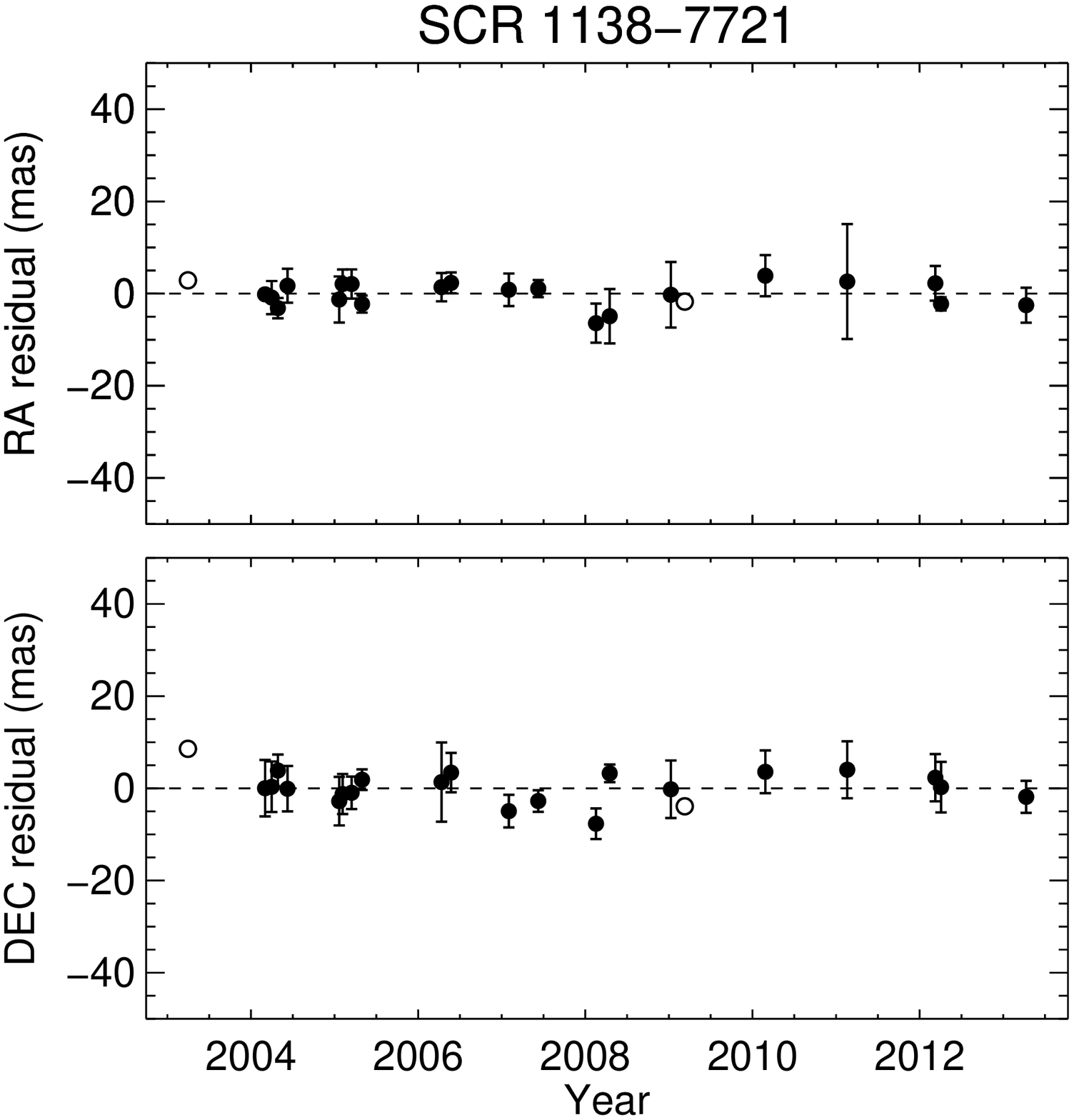}}
\hspace{-0.25in}
\subfigure
{\includegraphics[scale=0.225]
{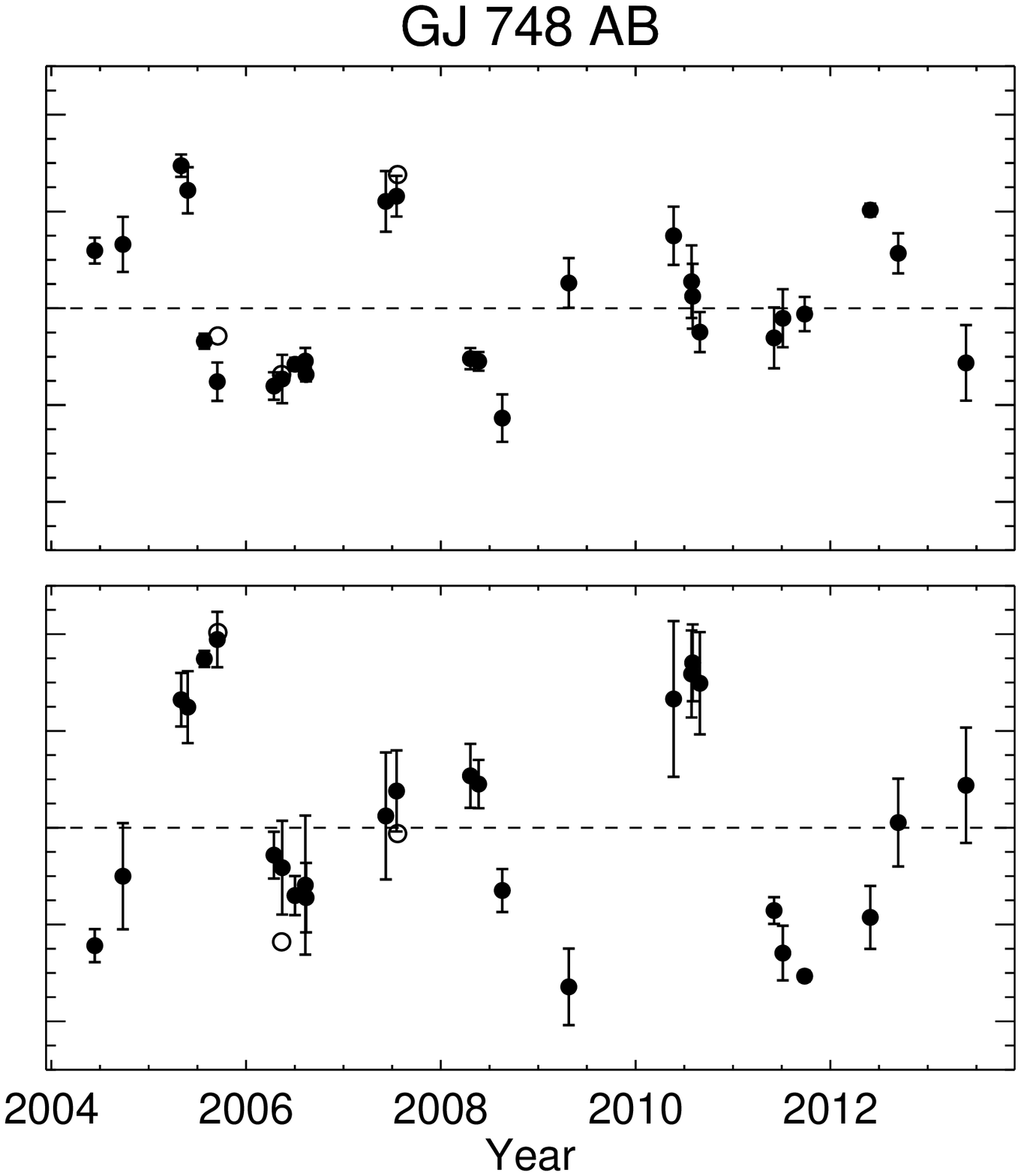}}
\hspace{-0.25in}
\subfigure
{\includegraphics[scale=0.225]
{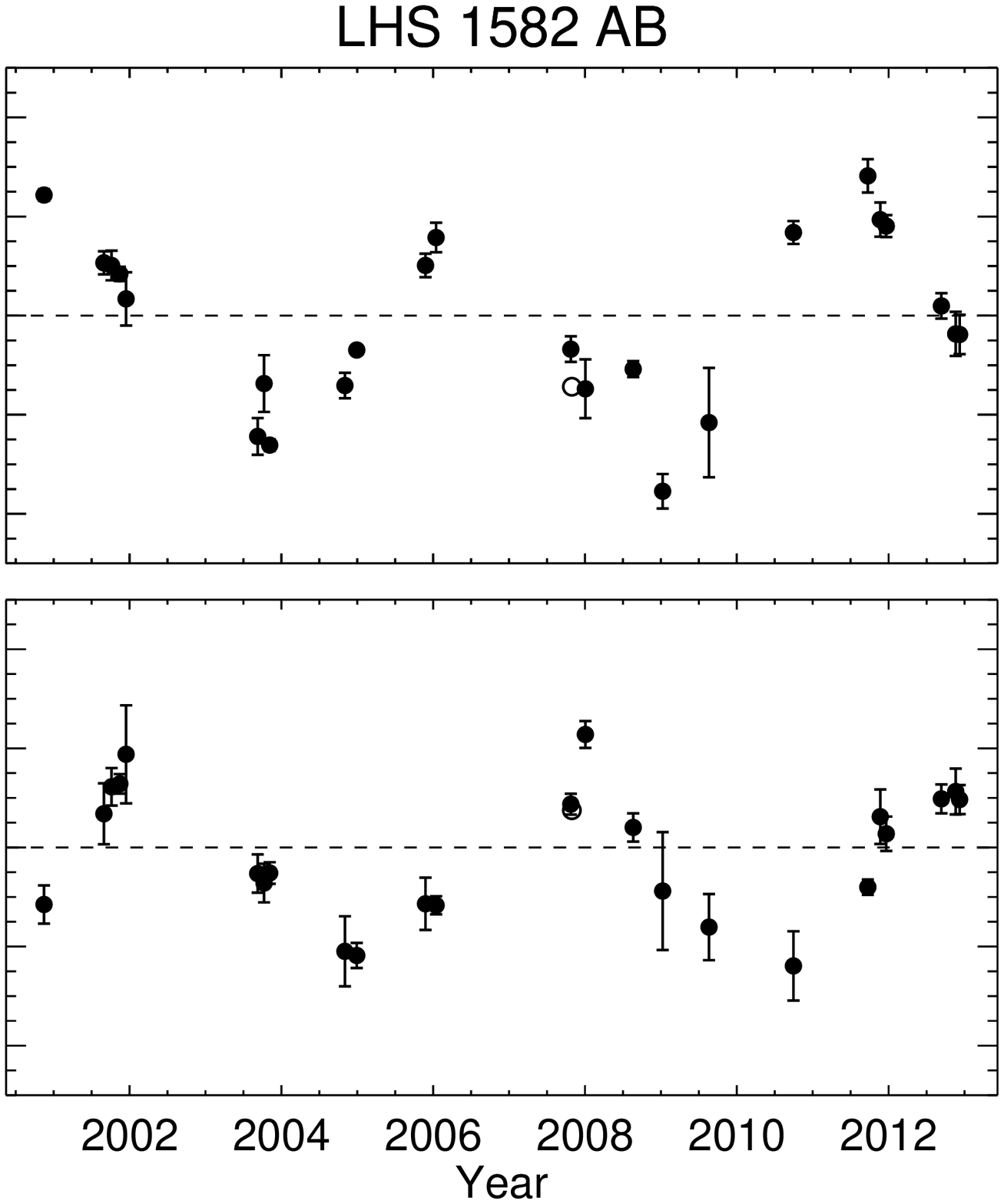}}
\hspace{-0.25in}
\subfigure
{\includegraphics[scale=0.225]
{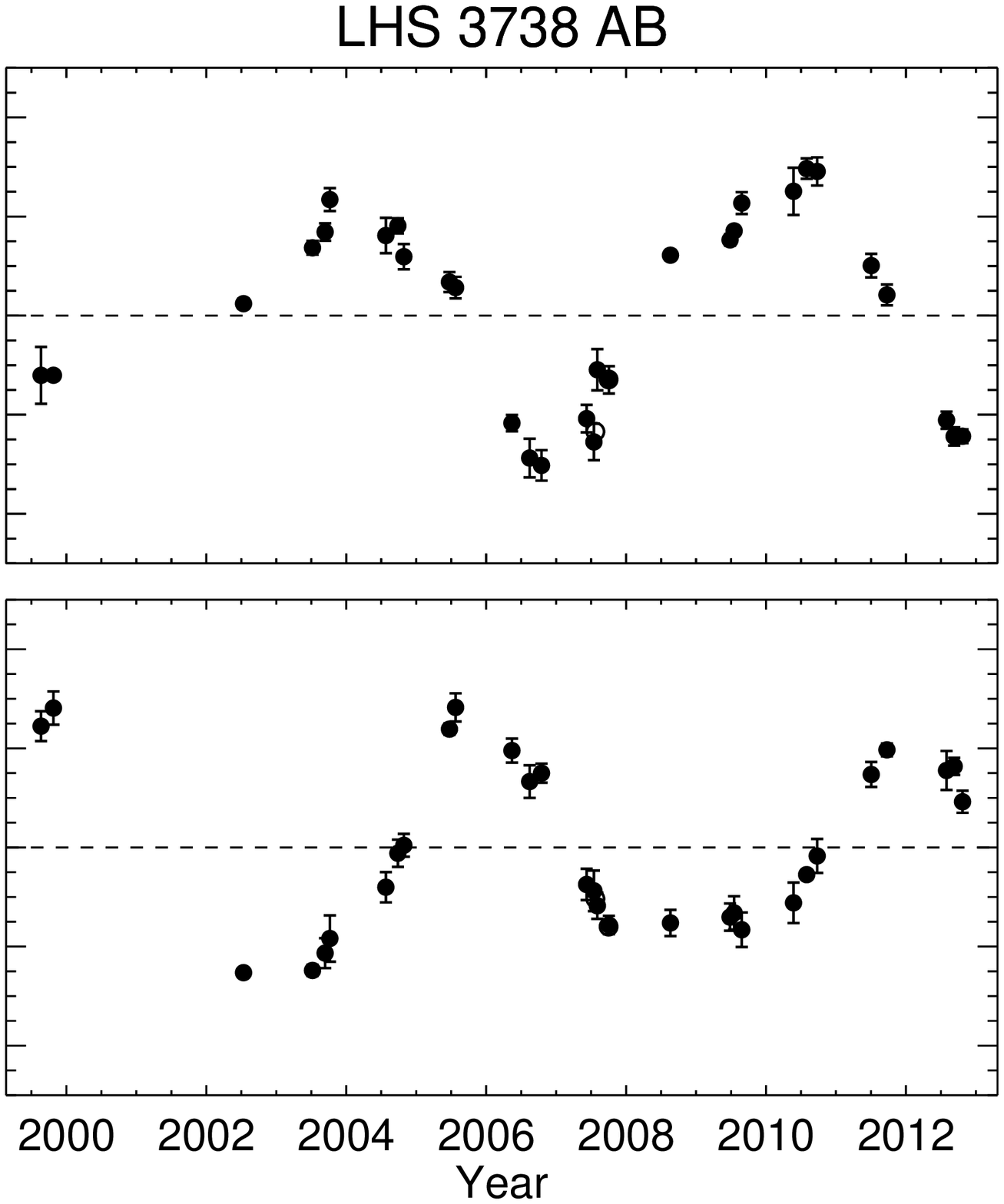}}

\caption{Astrometric residuals plotted in Right Ascension (R.A.) and
Declination (DEC), in units of milliarcseconds (mas). Filled circles
represent the mean of typically three to five frames taken in a single
night. Open circles represent nights for which there is only one
frame. All panels are on a $\pm$50 mas vertical scale.}
\label{fig1}
 
\end{figure*}

Astrometric residuals represent the deviation in a target's measured
position from the solution for proper motion and parallax, given for
all 16 target stars in Table \ref{tab1}. The residuals for
each star are plotted in Figure \ref{fig1}.  Each filled
circle represents the mean of typically three to five frames taken in
a single night, with a corresponding estimate of the nightly mean
error.  Open circles represent nights with only one frame, which are
included in the parallax measurement but are not included in the
following analysis of the residuals. The three binaries (GJ 748AB, LHS
1582AB, and LHS 3738AB) exhibit large, periodic perturbations in both
R.A. and Decl., indicating the presence of a companion in each
case. Because the systems are unresolved in our images, we calculate a
photocentric semi-major axis, as previously discussed. The remaining
targets generally have flat residuals, although some are more
scattered (GJ 849) than others (GJ 1128). Targets with the flattest
residuals, such as DENIS J1048-3956, generally have at least 8
reference stars that closely and evenly surround the target, and are
bright enough to have a least 1,000 peak counts when the target has
50,000. Fainter targets tend to have the flattest residuals because
they have a larger number of suitable reference stars and require
longer exposures, which smooth out short-term seeing effects that
produce PSFs with poorly-defined centroids.

\begin{figure*}[ht!]
\centering

{\includegraphics[scale=0.225]
{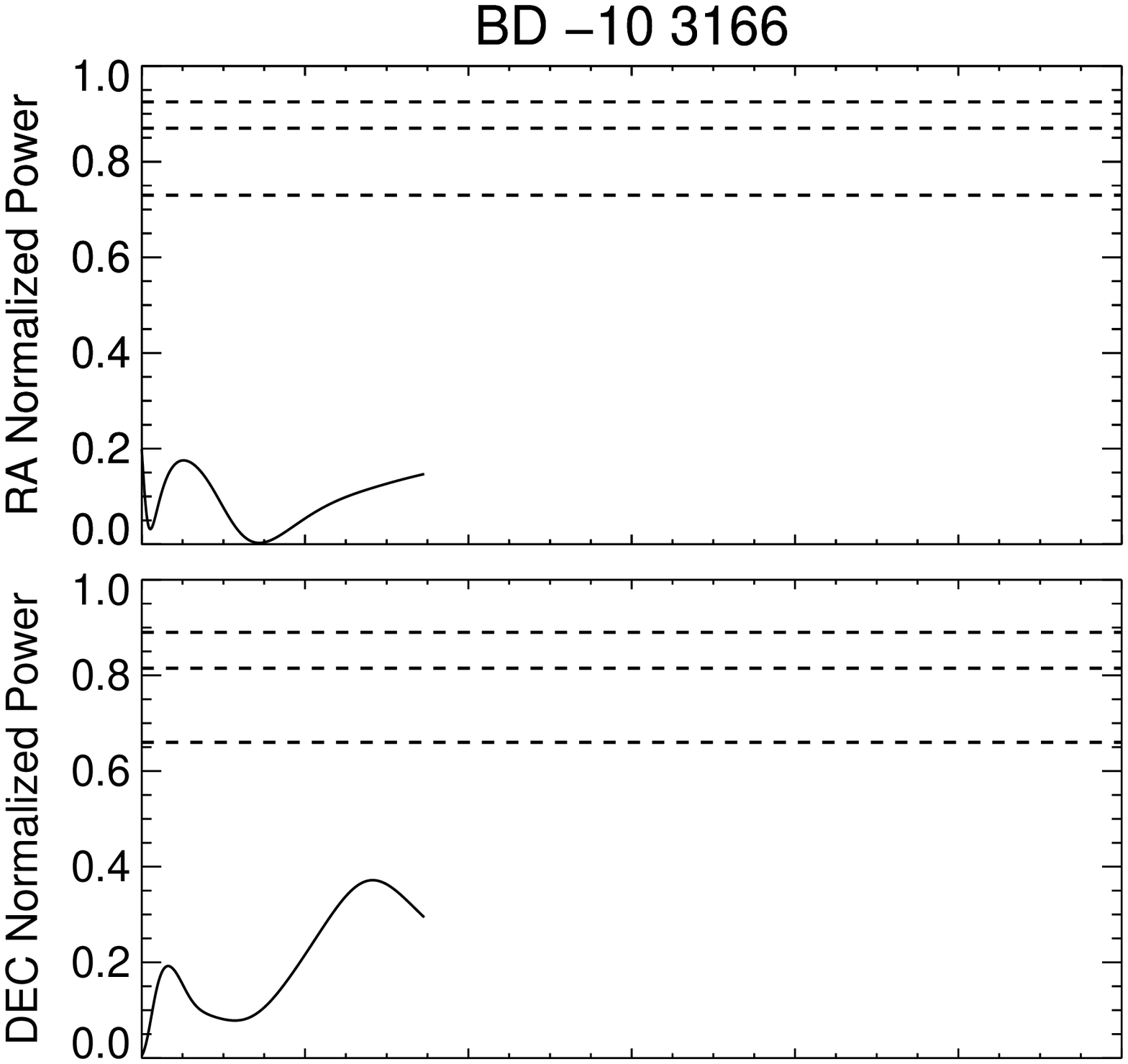}}
\hspace{-0.25in}
\subfigure
{\includegraphics[scale=0.225]
{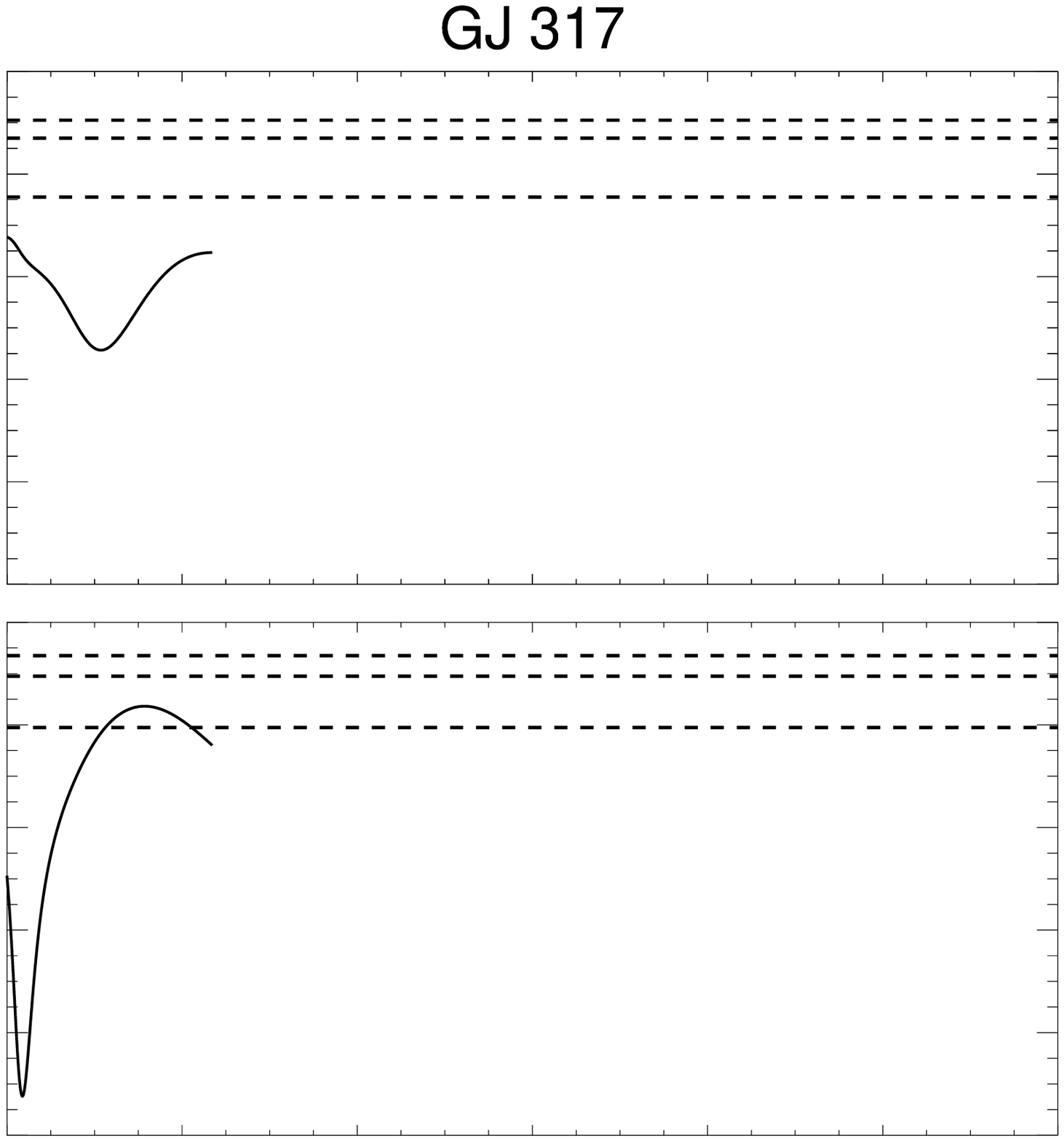}}
\hspace{-0.25in}
\subfigure
{\includegraphics[scale=0.225]
{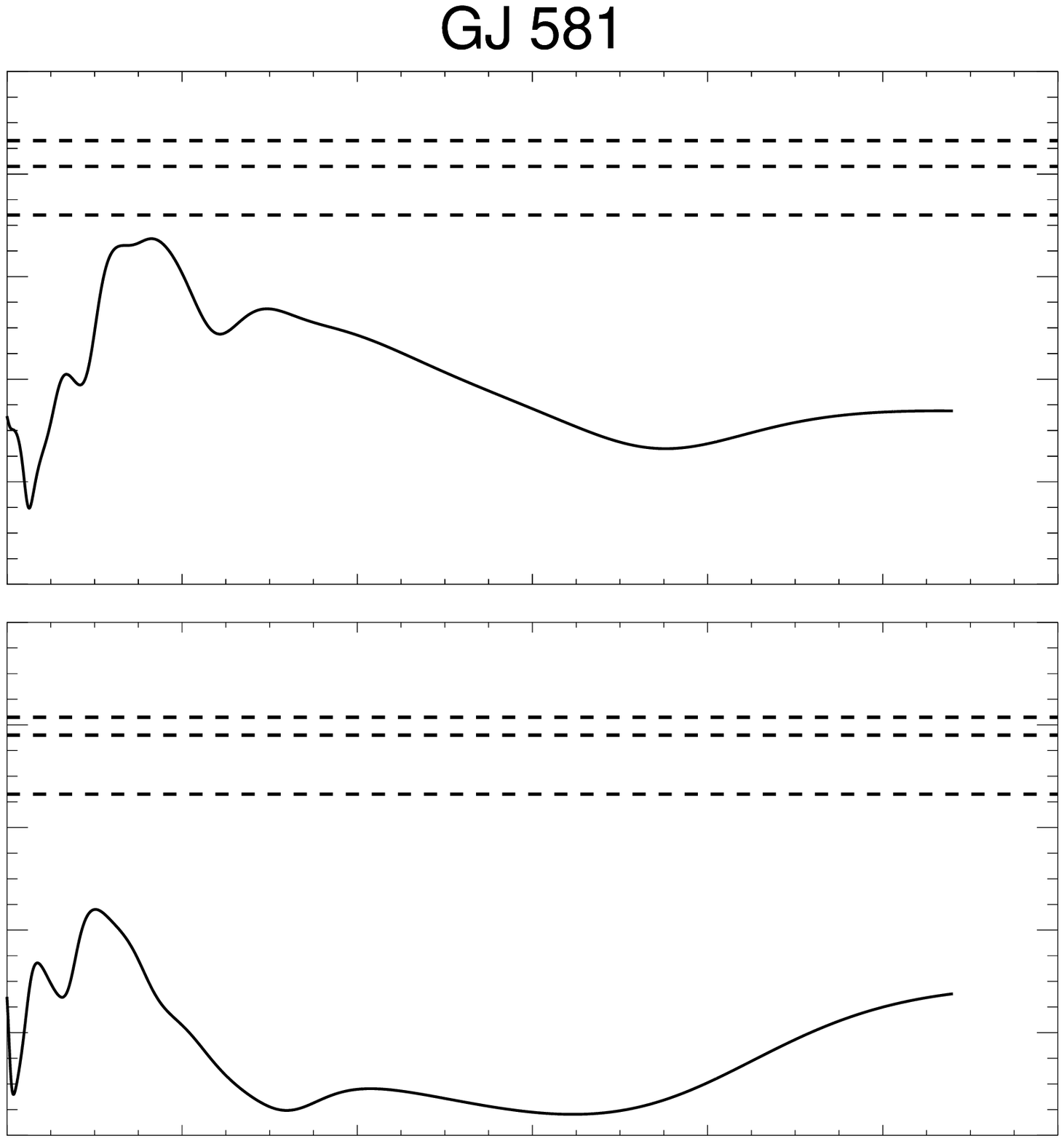}}
\hspace{-0.25in}
\subfigure
{\includegraphics[scale=0.225]
{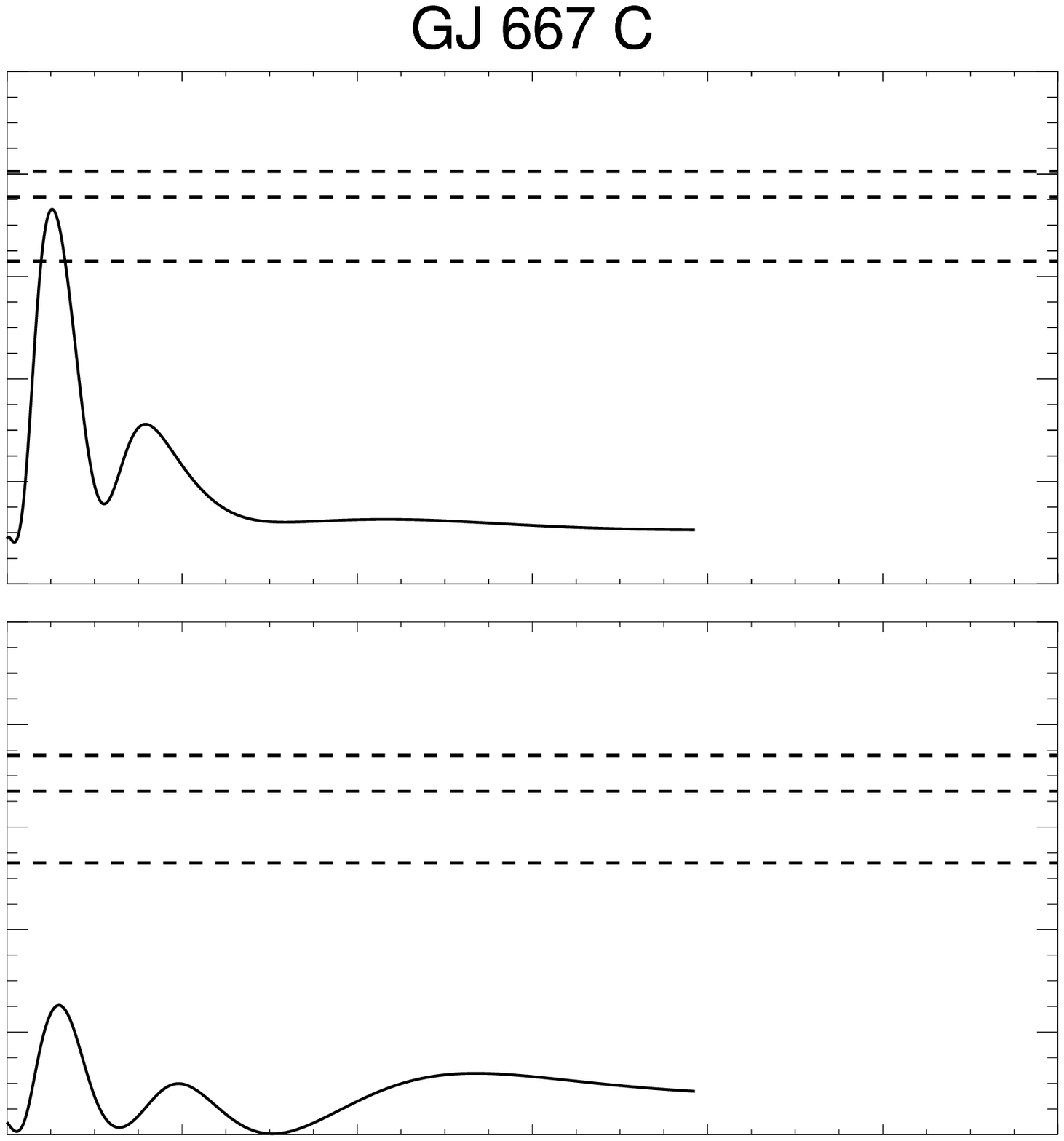}}


{\includegraphics[scale=0.225]
{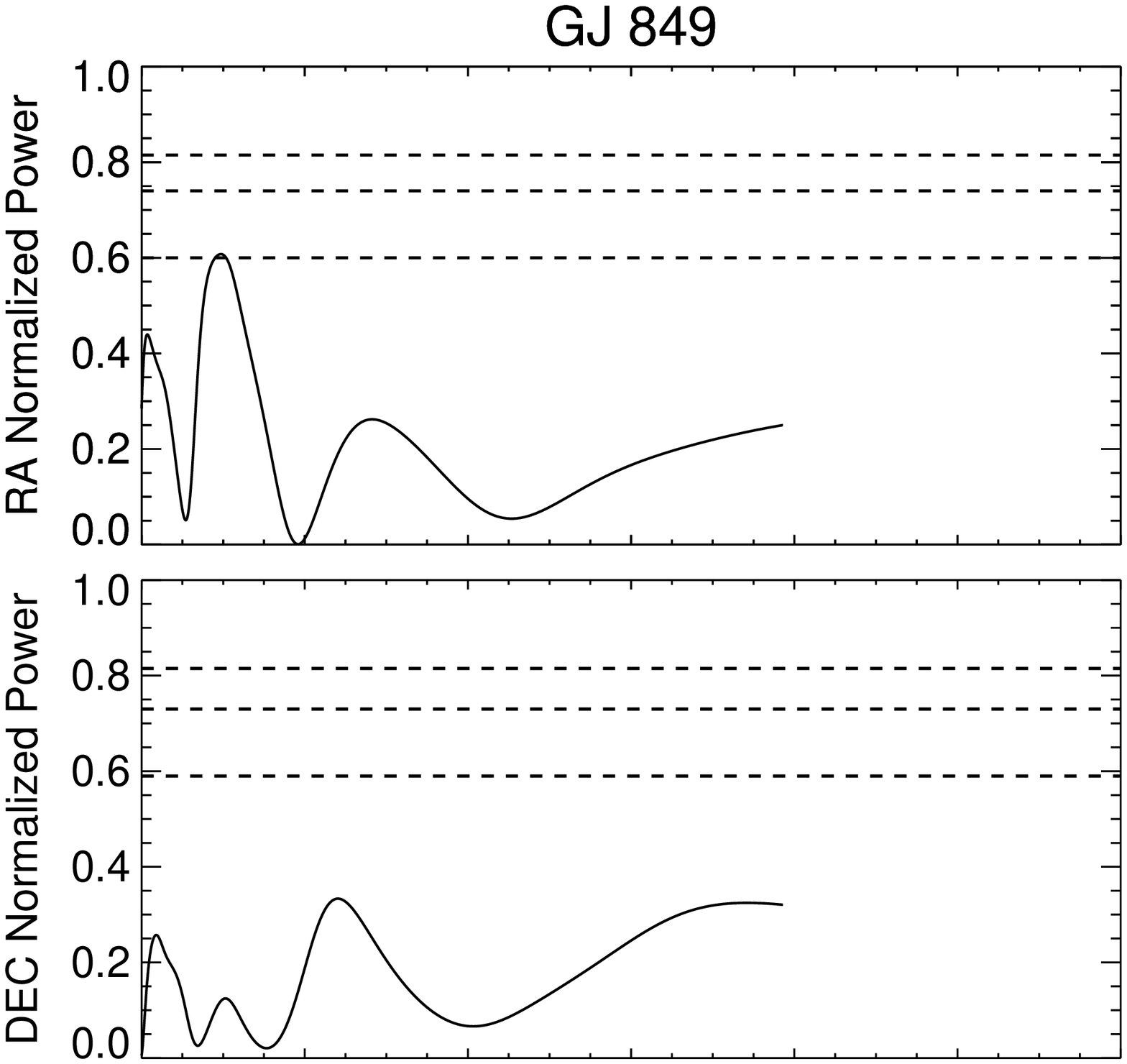}}
\hspace{-0.25in}
\subfigure
{\includegraphics[scale=0.225]
{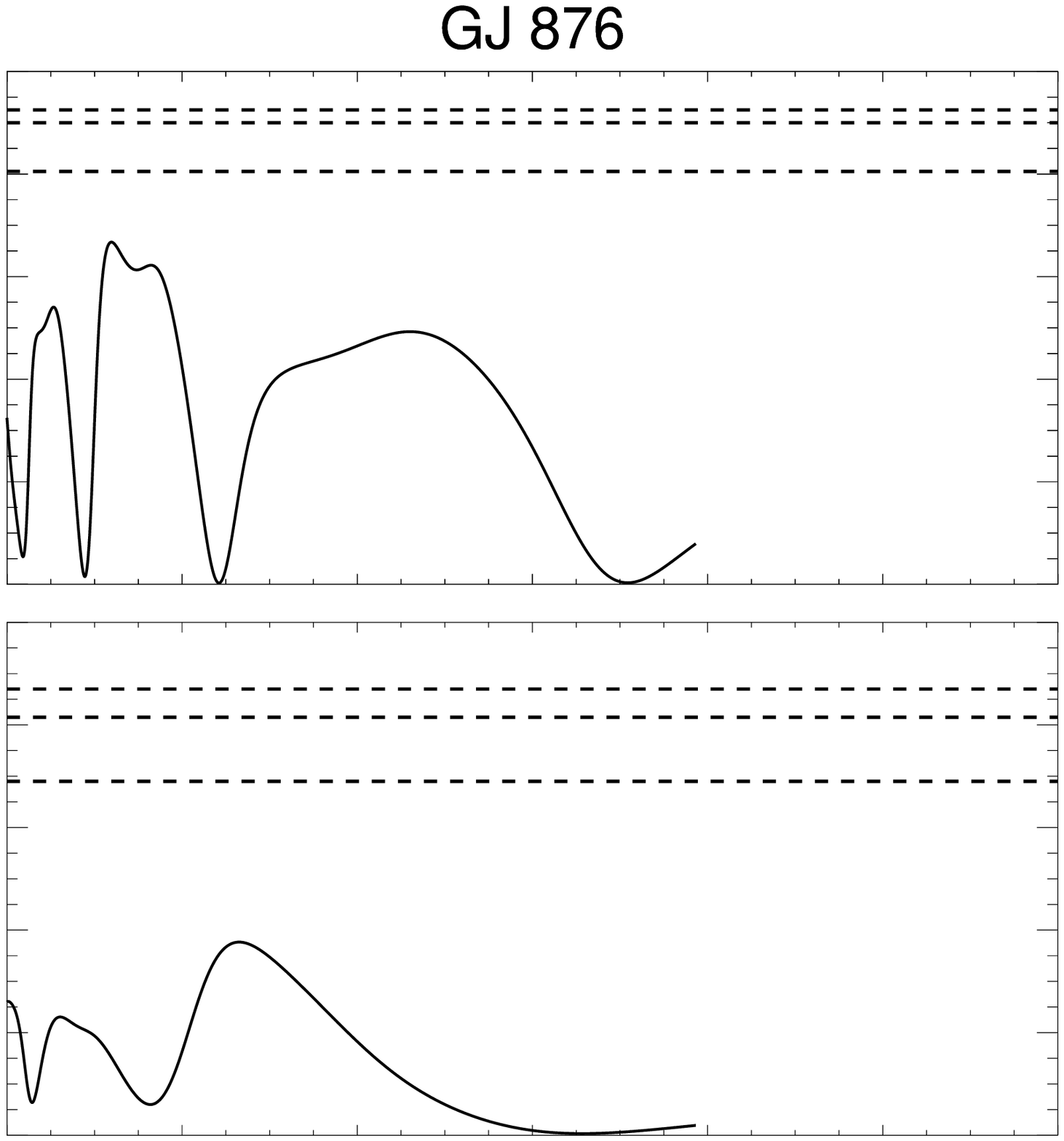}}
\hspace{-0.25in}
\subfigure
{\includegraphics[scale=0.225]
{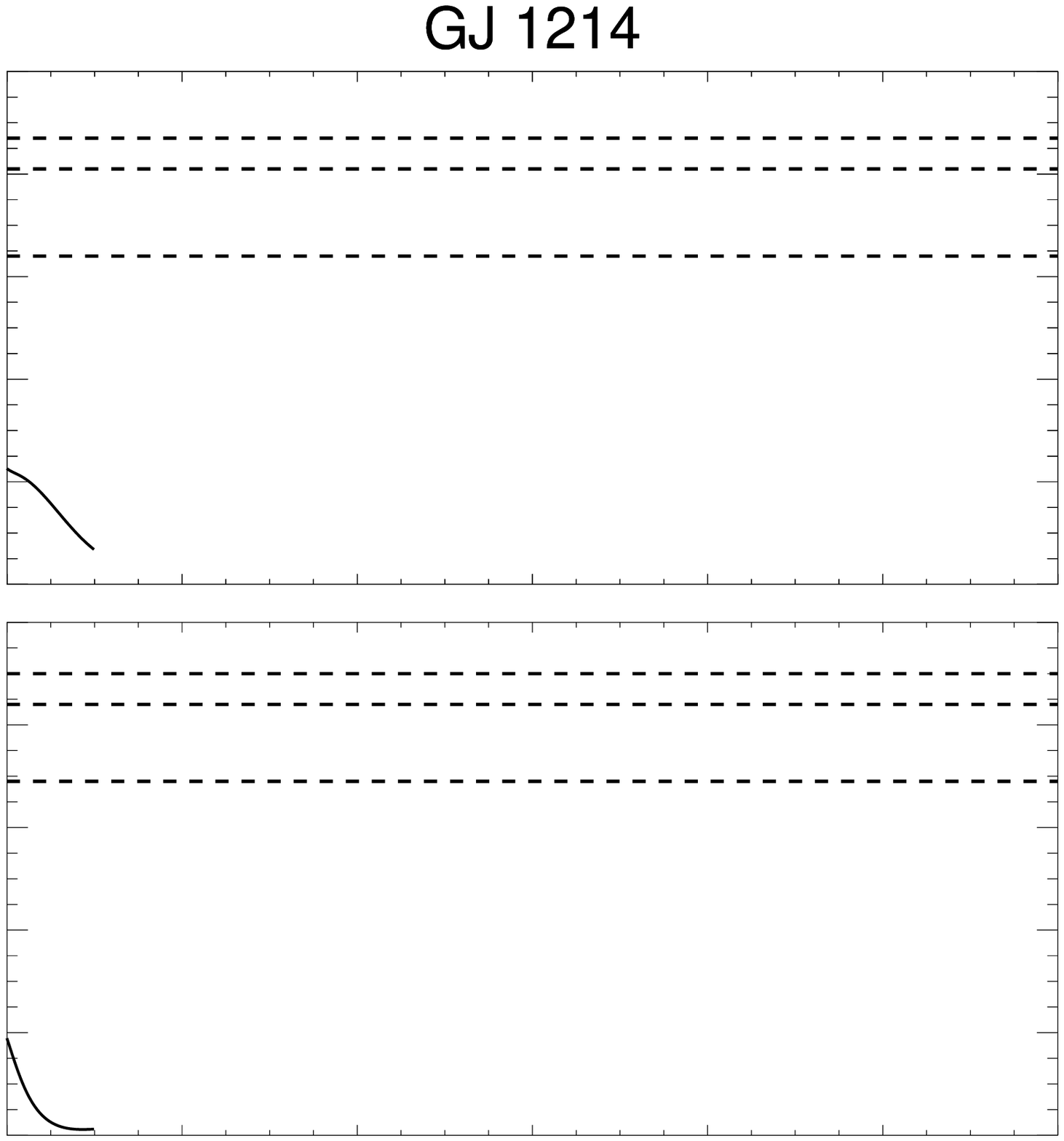}}
\hspace{-0.25in}
\subfigure
{\includegraphics[scale=0.225]
{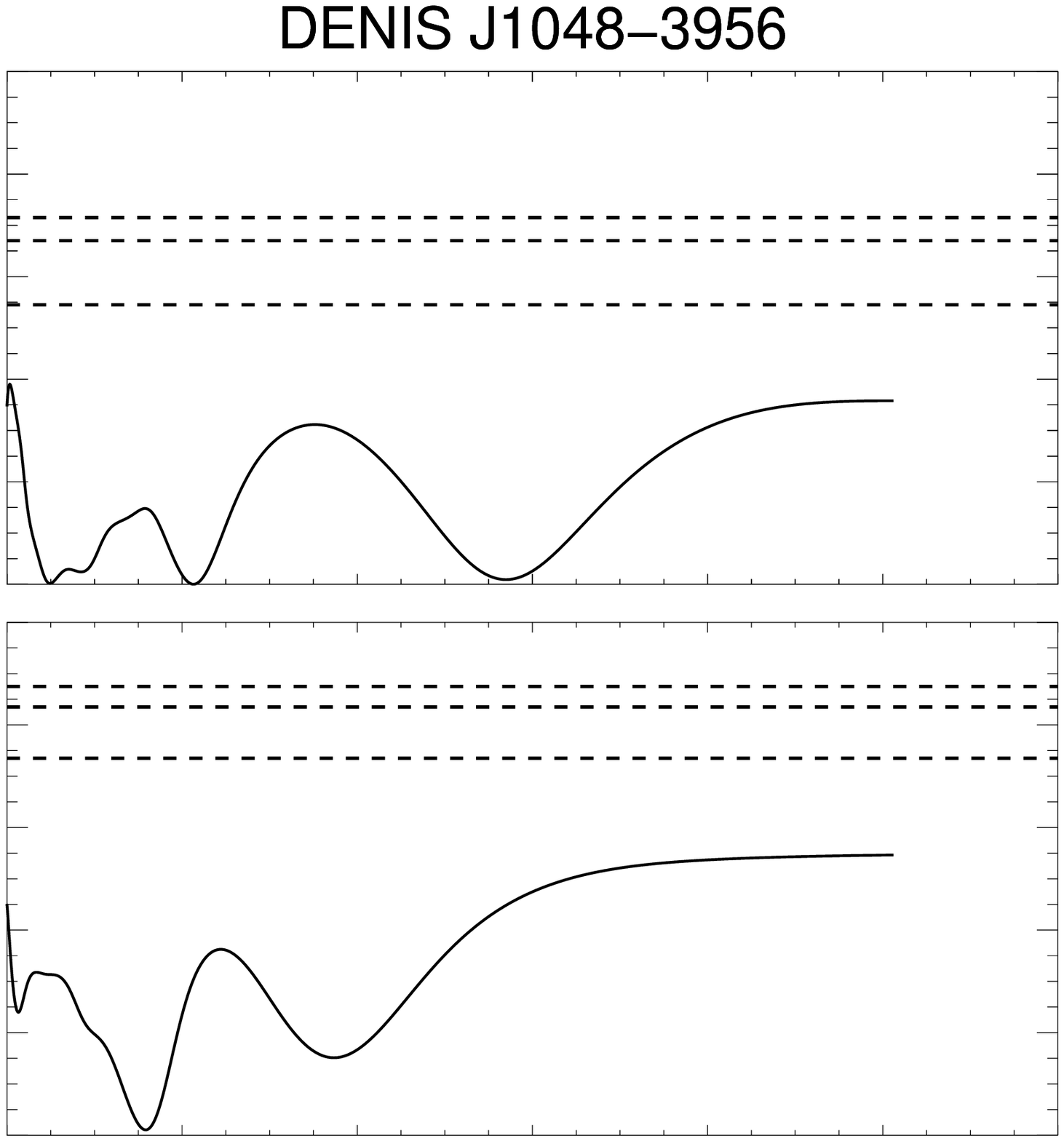}}


{\includegraphics[scale=0.225]
{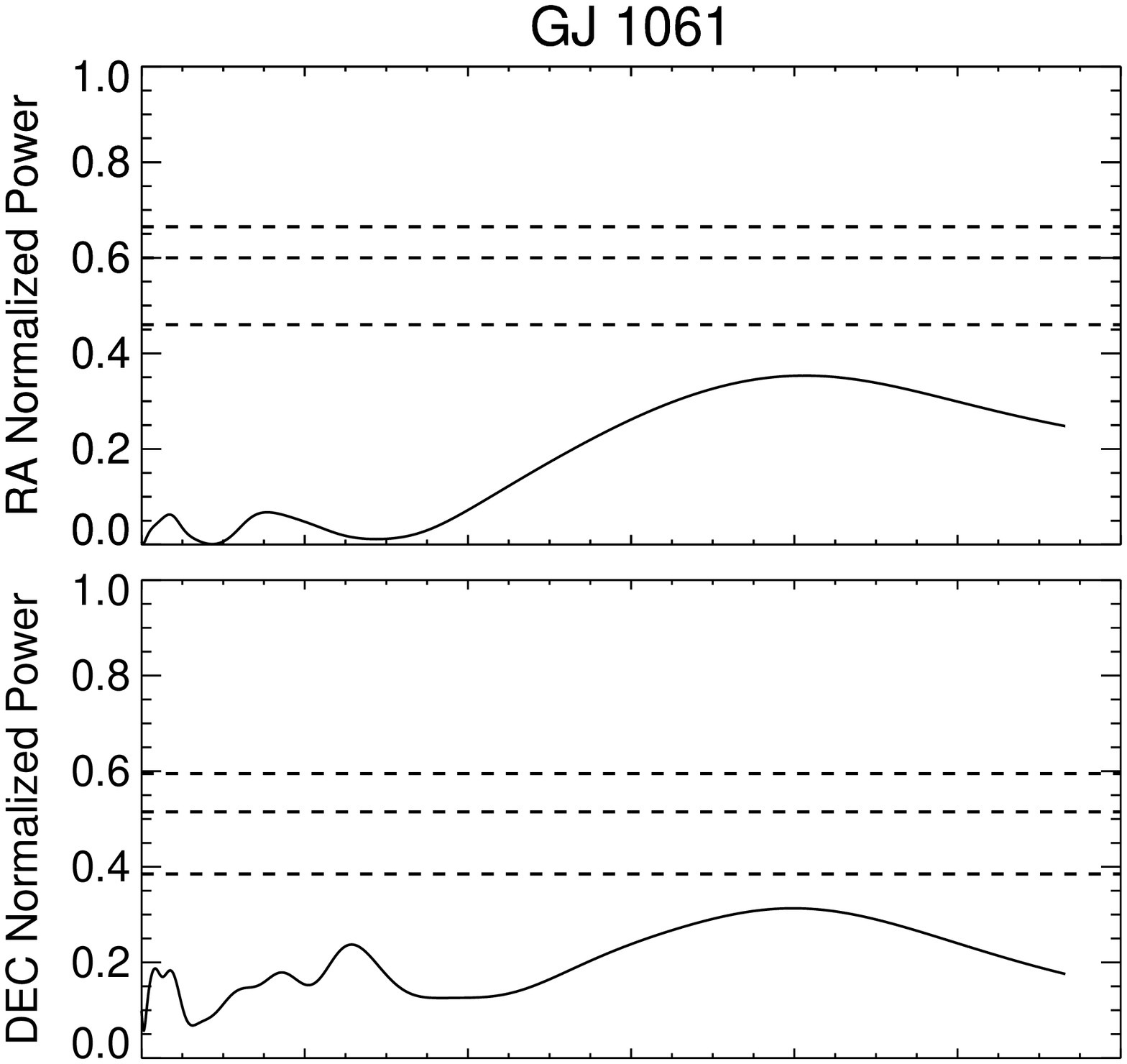}}
\hspace{-0.25in}
\subfigure
{\includegraphics[scale=0.225]
{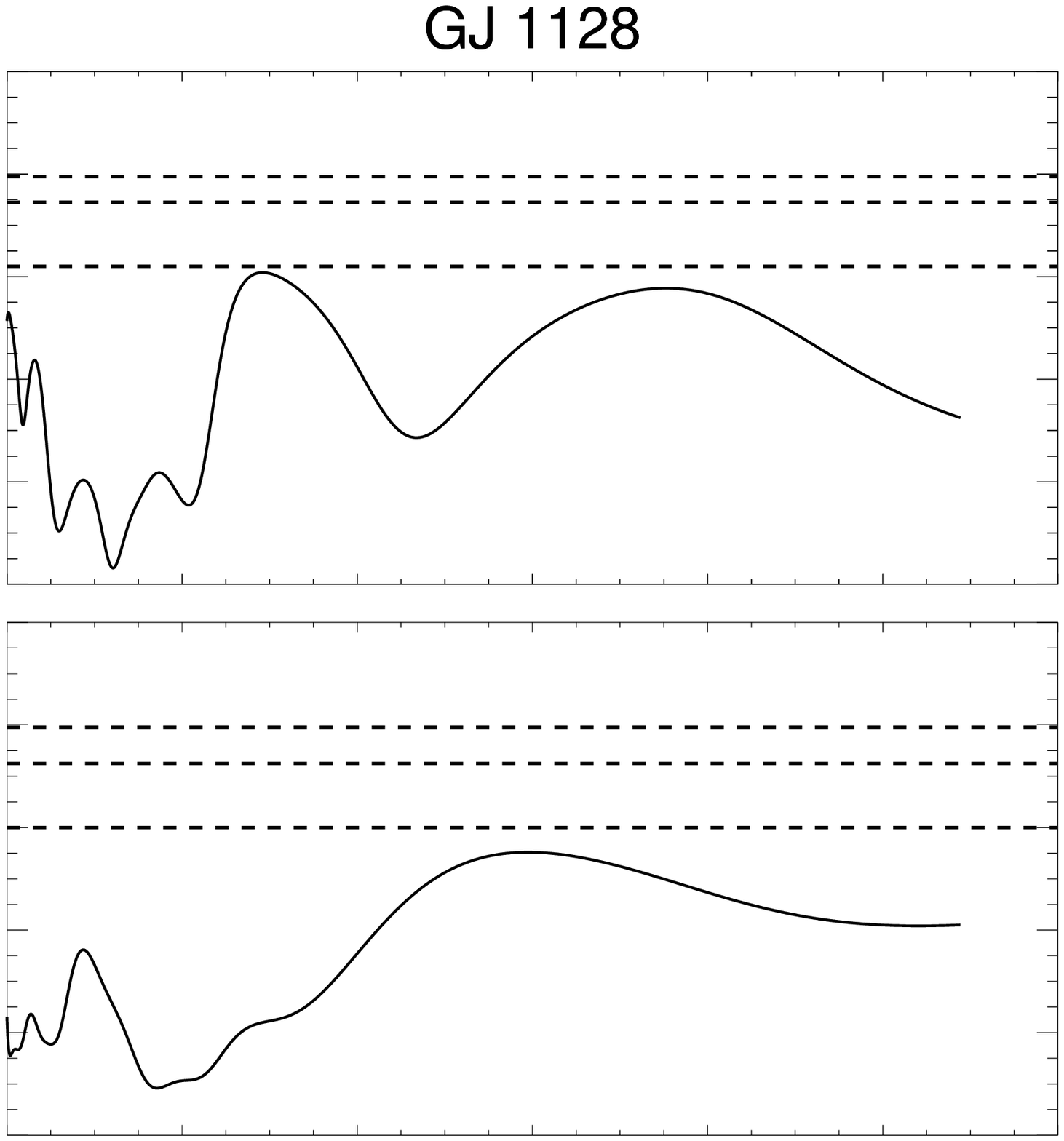}}
\hspace{-0.25in}
\subfigure
{\includegraphics[scale=0.225]
{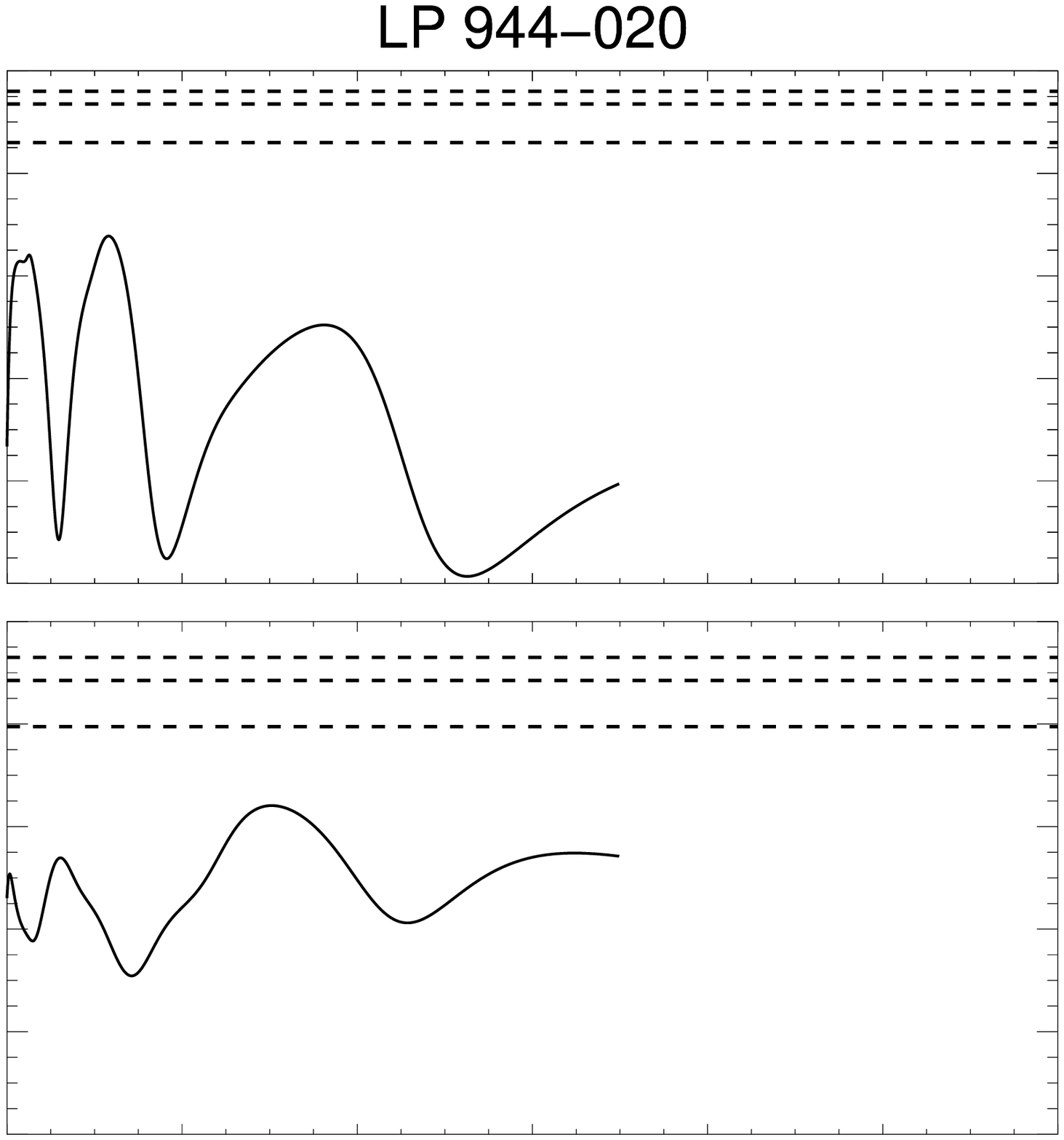}}
\hspace{-0.25in}
\subfigure
{\includegraphics[scale=0.225]
{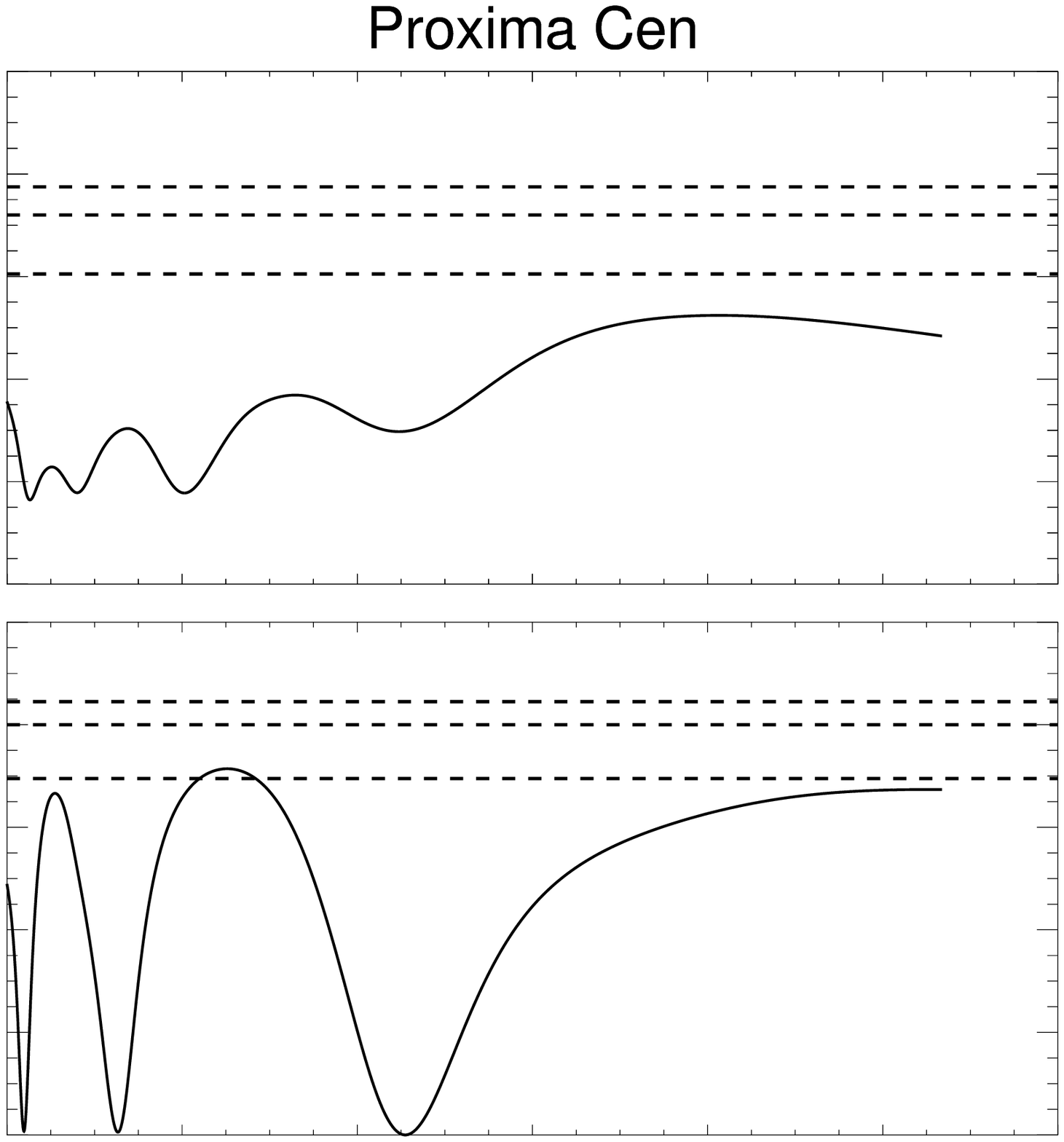}}


{\includegraphics[scale=0.225]
{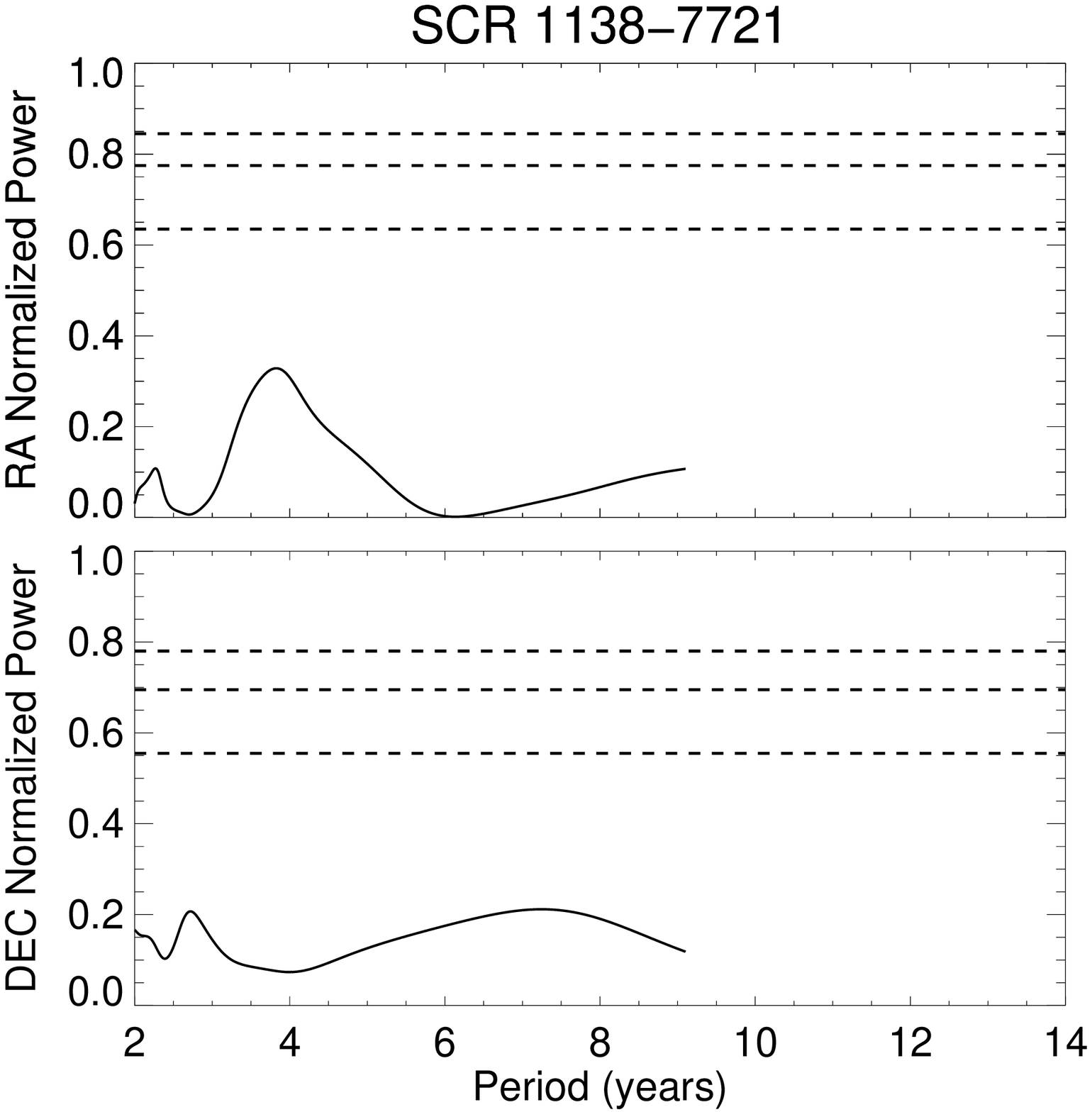}}
\hspace{-0.25in}
\subfigure
{\includegraphics[scale=0.225]
{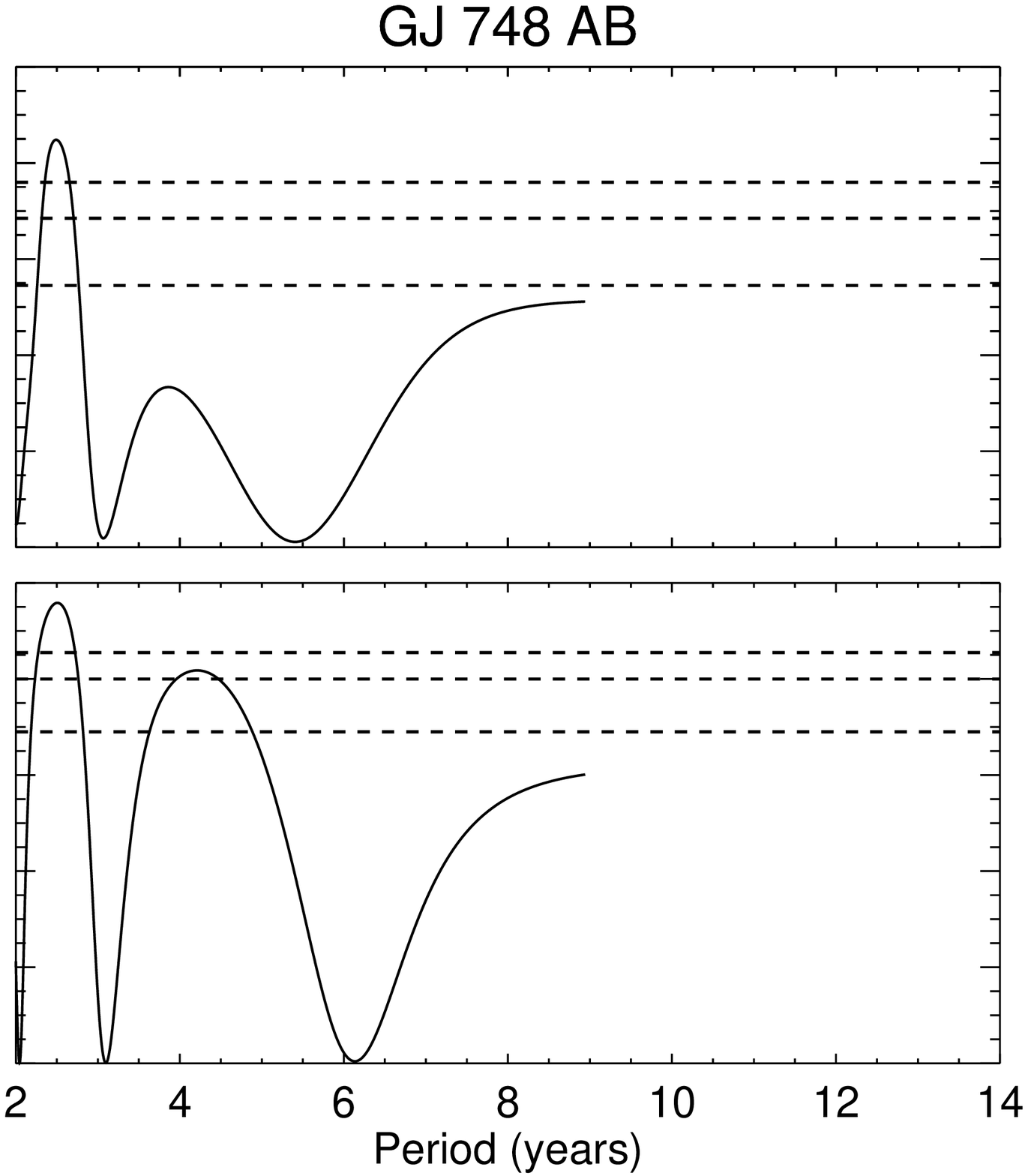}}
\hspace{-0.25in}
\subfigure
{\includegraphics[scale=0.225]
{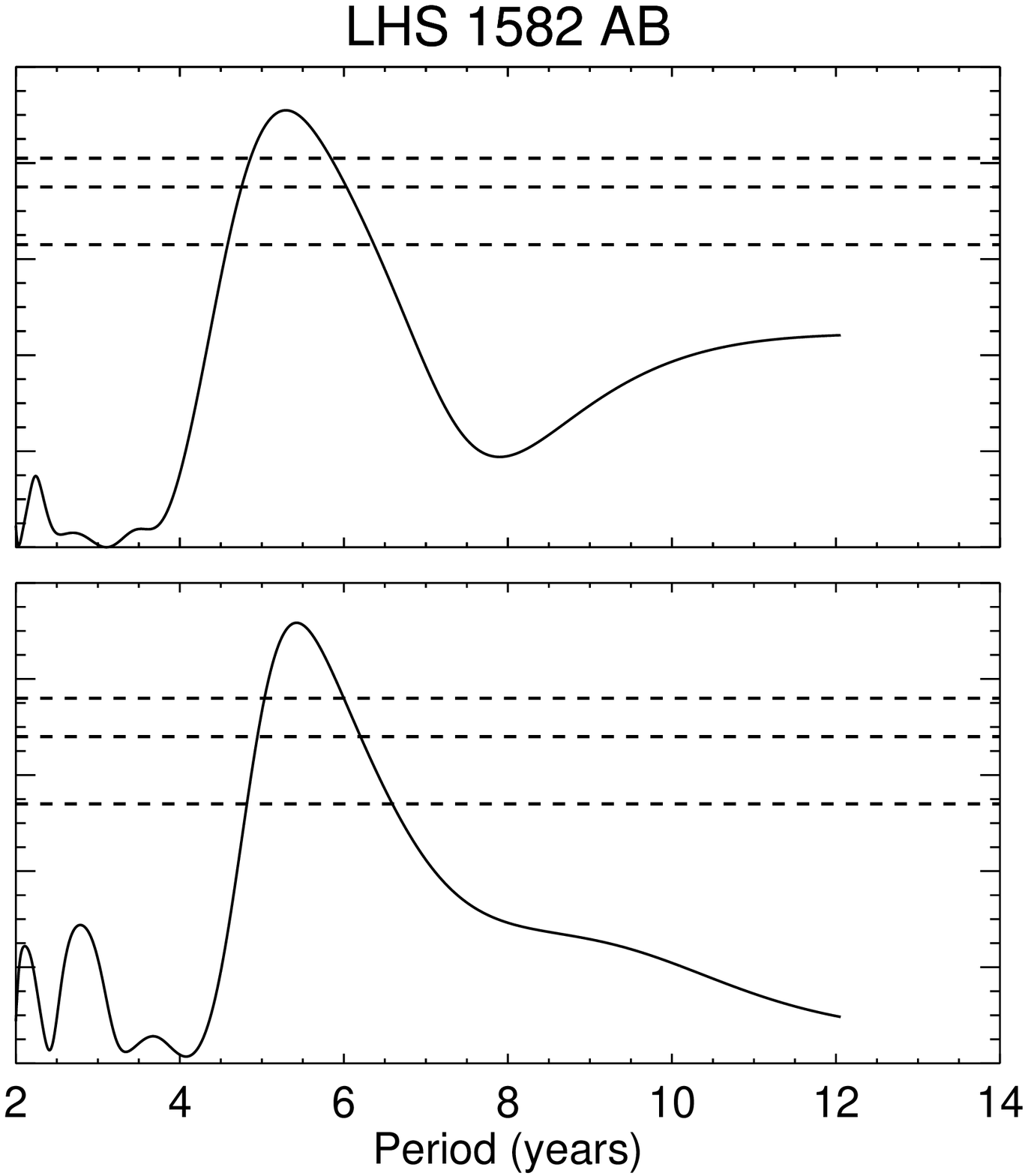}}
\hspace{-0.25in}
\subfigure
{\includegraphics[scale=0.225]
{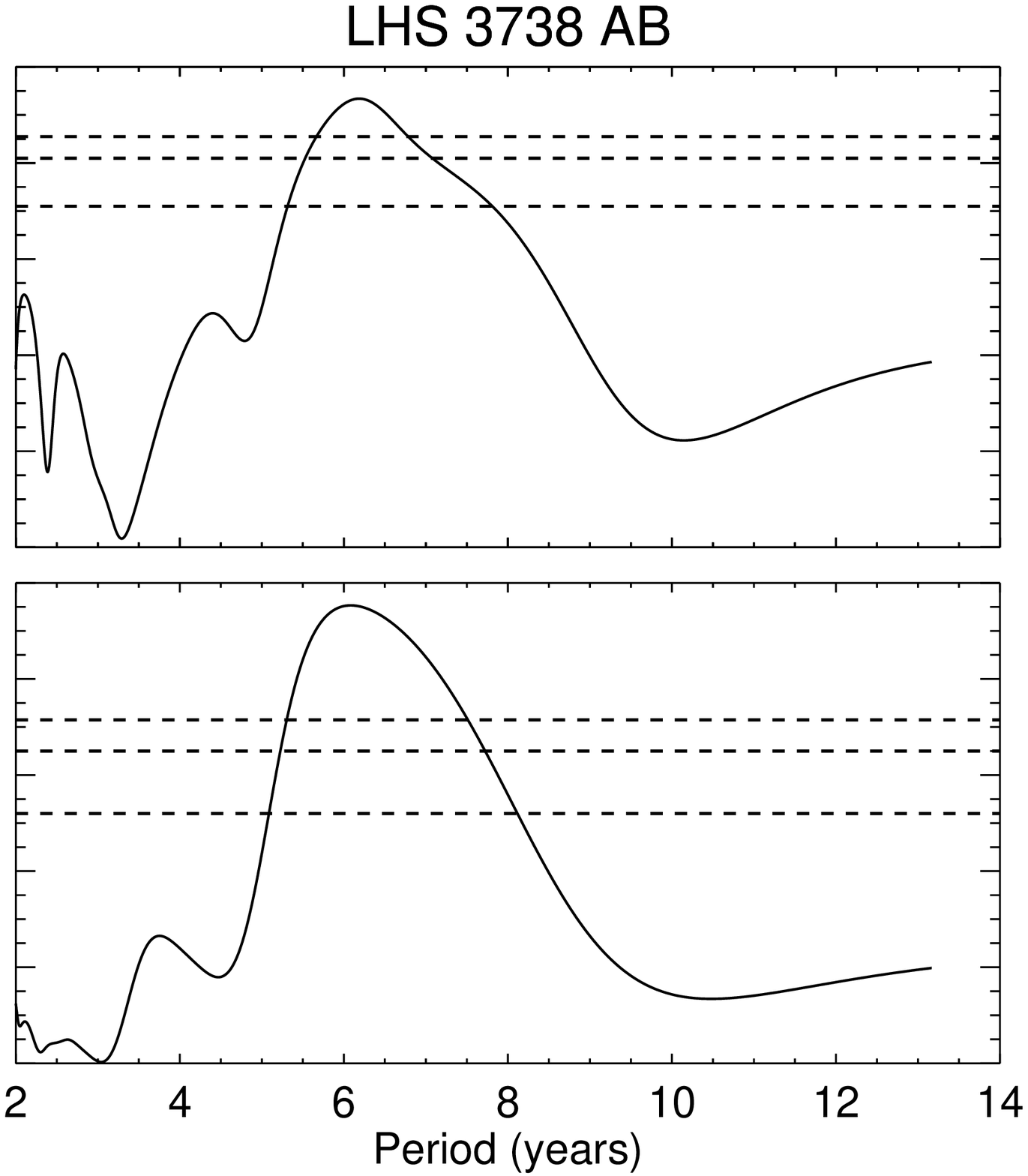}}

\caption{Periodograms and false alarm probabilities (FAP) for the
astrometric residuals. The dashed lines correspond to FAP values of
0.1 (lowest line), 0.01, and 0.001 (highest line). For the three
binaries, the periodograms have significant peaks (FAP $<$ 0.001) at
the same period in both R.A. and Decl., while none of the other targets
have any significant peaks even at the 0.01 level.}
\label{fig2}
 
\end{figure*}

The residuals were analyzed for the presence of companions using
Lomb-Scargle periodograms, shown in Figure \ref{fig2}, generated by
the method given in \citet{zec09}. This method weights each data point
by its estimated nightly mean error, and produces a normalized
periodogram for the data sequence with power ranging from 0 to 1. We
only attempt to detect companions with periods greater than two years
but less than the length of the observations, and have only generated
periodograms for those ranges. The data were oversampled in increments
of one day to create smooth periodograms. As \citet{fre07} have shown,
oversampling the periodogram does not significantly increase the risk
of false periodogram peaks.

As the data are irregularly spaced, it was not possible to use an
analytical formula for the false alarm probabilities (FAP). Instead,
empirical FAP distribution functions were calculated as outlined in
\citet{fre07} by generating periodograms for 10,000 sets of random
noise. Each of the 10,000 data sets has the same number of
observations as the actual data, with the same observation times and
errors as the actual data. The randomized values of each point within
a given data set are assigned from a normal distribution with a
standard deviation representing the distribution of offsets from zero
of the actual data points.

Each of these 10,000 periodograms was then sampled in a grid of
periods with five day increments between two years and the length of
observations to find the highest power occurring in each
periodogram. The FAP for an observed periodogram peak of a given power
is equal to the fraction of the 10,000 data sets with peaks greater
than or equal to that power. The normalized powers corresponding to
FAP values of 0.1, 0.01, and 0.001 are shown as dashed lines in Figure
\ref{fig2}. This corresponds to 10\%, 1\%, and 0.1\% probabilities
that the peak is due solely to random noise and the cadence of
observations. As a calibration, periodograms and FAP functions were
also generated for the three confirmed stellar binaries for which
astrometric perturbations are clearly evident. These binaries have
significant periodogram peaks (FAP $<$ 0.001) in both the R.A. and
Decl. axes at periods corresponding to their orbital periods, while
the remaining 13 targets have no significant peaks. This indicates
that our periodograms are sensitive to astrometric perturbations, and
that we have not detected any significant periodicity in the residuals
of the remaining 12 M dwarfs and one K dwarf without companions.

Given that there are no companions evident in the residuals of the
non-binary targets, we aim to establish lower limits for the companion
masses and periods to which we are sensitive, and would have detected
were they present. For each target we ran a simulation of 10 million
hypothetical companions orbiting each star with masses chosen randomly
from a uniform distribution between 0.5 and 80 $\mathrm{M_{Jup}}$, and
periods chosen randomly from a uniform distribution between 2 years up
to the length of observations. Although we have observed perturbations
as short as 1.2 years, our practical lower limit is 2 years. All
geometric parameters (inclination, eccentricity, time of periastron,
longitude of ascending node, and longitude of periastron) were
assigned randomly from a uniform distribution of all possible values,
including eccentricity, which was allowed to be as high as 0.99.

The program then calculates the astrometric perturbation that each
simulated companion would induce on the primary, using a primary mass
calculated from the mass-luminosity relations in \citet{hen93}
(equations 5a and 5b) and the revised relation for the lowest mass
stars in \citet{hen99}. For DEN J1048-3956 and LP 944-020, which are
too faint for the relations, we assume masses of 0.08
$\mathrm{M_{\Sun}}$. We assume that the companion does not contribute
significantly to the overall flux in each system, so that the
photocenter of the system is concentric with the primary star. This
assumption proves problematic when considering brown dwarf companions
to the two latest M dwarfs, DEN J1048-3956 and LP 944-020. However,
the trigonometric parallaxes of these targets agree well with their
photometric distance estimates (Columns 12 and 13 of Table
\ref{tab2}), and we conclude that neither have nearly equal
luminosity companions.

The goodness-of-fit between the observed data $y_i$ with errors
$\sigma_{i}$, and a flat line with $\tilde{y}_i = 0$ was determined
using the reduced chi squared statistic
\begin{equation}
\chi^{2}_{\mathrm{red}} = \frac{\chi^{2}}{K} = \frac{1}{K} \sum_{i=1}^{N} \left( \frac{y_{i} - \tilde{y}_{i}}{\sigma_{i}} \right)^{2},
\end{equation}
where $K$ is the number of degrees of freedom, given by $N - P - 1$
for $N$ data points and $P$ fitted parameters. Because we did not
attempt to fit a model to the data, but only to analyze how well a
simulation fits the data, we set $P$ equal to zero. As data points
with small error bars are more heavily weighted, we discarded epochs
that have unrealistic errors smaller than 1 mas --- representing less
than 5\% of the epochs --- because such points overconstrain the
orbits that can be fit.

\begin{figure}[ht!]
\centering

\includegraphics[trim = 1cm 0cm 2.25cm 0cm, clip, width=2.45in]{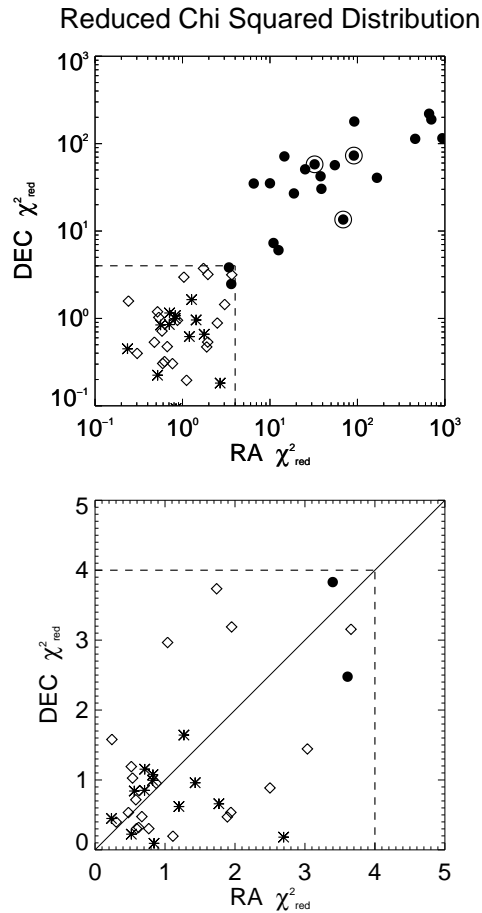}

\caption{Reduced chi squared ($\chi^{2}_{\mathrm{red}}$) values
  comparing the astrometric residuals to the case of perfectly flat
  residuals, i.e., an exact astrometric solution with no
  perturbation. The lower panel is a zoom of the dotted square region
  in the upper. Asterisks represent the 13 stars without close stellar
  secondaries analyzed in this paper. Diamonds represent additional
  calibration stars with flat residuals. Solid points represent binary
  stars with perturbations, including the three binaries analyzed in
  this paper, which are circled. All of the targets with confirmed
  perturbations have $\chi^{2}_{\mathrm{red}}$ values greater than 4
  in both R.A. and Decl., with the exception of two stars with
  long-term perturbations that have gaps in the astrometric
  observations.  All the targets without perturbations have
  $\chi^{2}_{\mathrm{red}}$ less than 4 in both axes.}
\label{fig3}

\end{figure}

Simulated orbits for which $\chi^{2}_{\mathrm{red}}$ was greater than
4 in at least one of either R. A. or Decl. we consider to be orbits
that we would have detected.  This threshold of 4 is based empirically
on the values plotted in Figure \ref{fig3}, where the lower
panel is an inset of the upper.  The plotted $\chi^{2}_{\mathrm{red}}$
values were calculated for (1) the stars included in this paper, (2)
calibration stars with flat residuals we use to monitor potential
fluctuations in equipment and our data reduction pipeline, and (3)
additional known binaries with perturbations.  We compare these
targets' residuals to the case of flat residuals, i.e., an exact
astrometric solution with no perturbation. As expected, the binaries
with perturbations have large $\chi^{2}_{\mathrm{red}}$ values,
indicating that a flat line is a poor fit to the data. The two solid
points inside the $\chi^{2}_{\mathrm{red}} = 4$ box are long term
perturbations with large gaps in the astrometric observations.  As
more data are collected, those two points will move to larger
$\chi^{2}_{\mathrm{red}}$ values outside the box, but we include the
points here for completeness. For those stars with no perturbations,
all have $\chi^{2}_{\mathrm{red}}$ less than 4 in both axes, and are
centered around 1. This indicates that we are accurately calculating
our measurement errors, and provides an empirical
$\chi^{2}_{\mathrm{red}}$ value of 4, above which we are sensitive to
perturbations.  

\section{Results}
\label{sec:results}

Table \ref{tab1} gives the parallax and proper motion results for the
16 systems, with details about the astrometric observations (filters
used, number of seasons observed, number of frames used in reductions,
time coverage, span of time, and the number of reference stars) and
results (relative parallaxes, parallax corrections, absolute
parallaxes, proper motions, position angles of the proper motions, and
the derived tangential velocities based on relative proper motions and
parallaxes).  All but two of the sixteen systems have parallax errors
of $\sim$2 mas or less. BD $-$10 3166 and GJ 876 have larger errors
due to combinations of faint reference stars and short
exposures. Corrections to absolute parallax are generally less than
$\sim$2 mas, so systematics in the corrections should not
significantly affect the results. Three targets (GJ 317, GJ 667C, and
BD $-$10 3166, ) have corrections of $\sim$3$\,$-$\,$4 mas due to
reddening of the reference stars, which skews their photometric
distance estimates. In these cases, we adopt a generic correction of
1.50 $\pm$ 0.50 mas. The per observation precision for each target is
listed in Column 16, representing the mean of the observation errors
in R.A. and Decl. The percentage of companions eliminated listed in
Column 17 is discussed in $\S$\ref{sec:discuss}.

\newpage
\addtolength{\voffset}{2in}
\begin{landscape}
\begin{deluxetable}{lccccccccrrrrrrccc}
\centering
\tabletypesize{\tiny}
\setlength{\tabcolsep}{0.03in}
\tablewidth{0pt}
\tablecaption{Astrometric Results}
\tablehead{\colhead{}                    &
	   \colhead{R.A.}                &
 	   \colhead{Decl.}               &
 	   \colhead{}                    &
	   \colhead{}                    &
	   \colhead{}                    &
	   \colhead{}                    &
	   \colhead{}                    &
	   \colhead{}                    &
	   \colhead{$\pi$(rel)}          &
	   \colhead{$\pi$(corr)}         &
	   \colhead{$\pi$(abs)}          &
	   \colhead{$\mu$}               &
	   \colhead{P.A.}                &
	   \colhead{V${_\mathrm{tan}}$}   &
	   \colhead{Obs. Prec.}          &
	   \colhead{Perc.}               &
	   \colhead{}                    \\
           \colhead{Name}                &
	   \colhead{(J2000.0)}           &
 	   \colhead{(J2000.0)}           &
 	   \colhead{Fil.}                &
	   \colhead{N$_{\mathrm{sea}}$}  &
	   \colhead{N$_{\mathrm{frm}}$}  &
	   \colhead{Coverage}            &
	   \colhead{Years}               &
 	   \colhead{N$_{\mathrm{ref}}$}  &
	   \colhead{(mas)}               &
	   \colhead{(mas)}               &
	   \colhead{(mas)}               &
	   \colhead{(mas yr$^{-1}$)}     &
	   \colhead{(deg)}               &
	   \colhead{(km/s)}              &
	   \colhead{(mas)}               &
	   \colhead{Cmpn.}               &
	   \colhead{Notes}               \\
           \colhead{(1)}                 &
           \colhead{(2)}                 &
           \colhead{(3)}                 &
           \colhead{(4)}                 &
           \colhead{(5)}                 &
           \colhead{(6)}                 &
           \colhead{(7)}                 &
           \colhead{(8)}                 &
           \colhead{(9)}                 &
           \colhead{(10)}                &
           \colhead{(11)}                &
           \colhead{(12)}                &
           \colhead{(13)}                &
           \colhead{(14)}                &
           \colhead{(15)}                &
           \colhead{(16)}                &
           \colhead{(17)}                &
           \colhead{(18)}                }

\startdata
\multicolumn{17}{c}{Extrasolar Planet Hosts}\\
\hline
GJ 317           &  08 40 59.21 & $-$23 27 22.6 & R &   5c &   75 & 2009.04-2013.38 &  4.35 &   7 &    64.04 $\pm$ 1.45 & 1.50 $\pm$ 0.50 &  65.54 $\pm$ 1.53 &  930.7 $\pm$ 1.1 & 330.5 $\pm$ 0.13 & 65.5 &  4.99 &      &   \\   
BD $-$10  3166   &  10 58 28.79 & $-$10 46 13.4 & I &   7s &   71 & 2004.43-2011.50 &  7.07 &   6 &    13.84 $\pm$ 3.04 & 1.50 $\pm$ 0.50 &  15.34 $\pm$ 3.08 &  185.9 $\pm$ 1.5 & 269.1 $\pm$ 0.67 & 52.4 & 10.85 &      &   \\  
GJ 581           &  15 19 26.83 & $-$07 43 20.1 & V &  14s &  267 & 2000.58-2013.38 & 12.80 &  11 &   157.67 $\pm$ 1.57 & 1.12 $\pm$ 0.17 & 158.79 $\pm$ 1.58 & 1224.3 $\pm$ 0.4 & 266.0 $\pm$ 0.03 & 36.5 &  7.93 & 96\% & a \\  
GJ 1214          &  17 15 18.92 & $+$04 57 50.1 & I &   4c &   80 & 2010.39-2013.38 &  3.00 &   9 &    68.20 $\pm$ 1.26 & 1.88 $\pm$ 0.18 &  70.08 $\pm$ 1.27 &  945.5 $\pm$ 1.4 & 142.0 $\pm$ 0.17 & 63.9 &  5.02 &      &   \\  
GJ 667C          &  17 18 58.82 & $-$34 59 48.6 & V &  11s &  140 & 2003.52-2013.38 &  9.86 &   5 &   139.38 $\pm$ 1.98 & 1.50 $\pm$ 0.50 & 140.88 $\pm$ 2.04 & 1154.1 $\pm$ 0.6 & 101.0 $\pm$ 0.05 & 38.8 &  7.66 & 93\% &   \\  
GJ 849           &  22 09 40.34 & $-$04 38 26.8 & V &  11s &  135 & 2003.52-2013.39 &  9.86 &   5 &   113.78 $\pm$ 1.97 & 2.27 $\pm$ 0.30 & 116.05 $\pm$ 1.99 & 1118.0 $\pm$ 0.5 &  90.8 $\pm$ 0.04 & 45.7 &  9.53 & 81\% &   \\  
GJ 876           &  22 53 16.75 & $-$14 15 49.2 & V &  11s &   85 & 2003.52-2013.39 &  9.87 &   6 &   210.97 $\pm$ 3.99 & 2.14 $\pm$ 0.57 & 213.11 $\pm$ 4.03 & 1149.4 $\pm$ 1.1 & 125.7 $\pm$ 0.11 & 25.6 &  8.43 & 99\% &   \\    
\hline																									     
\multicolumn{17}{c}{Best Case Targets}\\																				     
\hline																									     
GJ 1061          &  03 35 59.72 & $-$44 30 45.5 & R &  13s &  194 & 1999.62-2012.95 & 13.32 &   7 &   269.92 $\pm$ 1.29 & 0.94 $\pm$ 0.08 & 270.86 $\pm$ 1.29 &  827.7 $\pm$ 0.3 & 117.7 $\pm$ 0.04 & 14.5 &  7.59 & 79\% & a \\
LP 944-020       &  03 39 35.25 & $-$35 25 43.8 & I &   8s &   59 & 2003.95-2012.94 &  8.99 &  10 &   154.53 $\pm$ 1.03 & 1.36 $\pm$ 0.10 & 155.89 $\pm$ 1.03 &  408.3 $\pm$ 0.3 &  48.5 $\pm$ 0.07 & 12.4 &  2.13 & 94\% & a \\
GJ 1128          &  09 42 46.36 & $-$68 53 06.1 & V &  13s &  167 & 2000.23-2013.12 & 12.89 &   8 &   153.54 $\pm$ 0.75 & 0.73 $\pm$ 0.11 & 154.27 $\pm$ 0.76 & 1123.0 $\pm$ 0.2 & 356.1 $\pm$ 0.02 & 34.5 &  3.17 & 80\% & a \\
DENIS J1048-3956 &  10 48 14.56 & $-$39 56 07.0 & I &  13s &  200 & 2001.15-2013.27 & 12.13 &  11 &   247.23 $\pm$ 0.60 & 0.85 $\pm$ 0.10 & 248.08 $\pm$ 0.61 & 1531.6 $\pm$ 0.2 & 229.5 $\pm$ 0.01 & 29.3 &  2.92 & 97\% & a \\
SCR 1138-7721    &  11 38 16.76 & $-$77 21 48.5 & I &  11s &  134 & 2003.25-2013.27 & 10.03 &  12 &   119.60 $\pm$ 1.01 & 0.81 $\pm$ 0.07 & 120.41 $\pm$ 1.01 & 2143.3 $\pm$ 0.4 & 287.8 $\pm$ 0.02 & 84.4 &  4.20 & 69\% & a \\
Proxima Cen      &  14 29 43.02 & $-$62 40 46.7 & V &  14s &  205 & 2000.57-2013.25 & 12.68 &   5 &   766.41 $\pm$ 0.91 & 1.72 $\pm$ 0.50 & 768.13 $\pm$ 1.04 & 3850.8 $\pm$ 0.6 & 282.4 $\pm$ 0.02 & 23.8 &  4.83 & 99\% & a \\
\hline																									    
\multicolumn{17}{c}{Confirmed Binaries}\\																				    
\hline																									    
LHS 1582AB      &  03 43 22.08 & $-$09 33 50.9 & R &  11s &  102 & 2000.87-2012.94 & 12.06 &   7 &    48.84 $\pm$ 1.18 & 2.00 $\pm$ 0.26 &  50.84 $\pm$ 1.21 &  509.4 $\pm$ 0.3 &  52.7 $\pm$ 0.06 & 47.5  &  3.88 &     & a \\
GJ 748AB        &  19 12 14.60 & $+$02 53 11.0 & V &  10s &  154 & 2004.45-2013.39 &  8.95 &  11 &    97.77 $\pm$ 1.15 & 2.22 $\pm$ 0.41 &  99.99 $\pm$ 1.22 & 1857.8 $\pm$ 0.5 & 107.4 $\pm$ 0.02 & 88.1  &  5.80 &     &   \\
LHS 3738AB      &  21 58 49.13 & $-$32 26 25.5 & R &  12s &  151 & 1999.64-2012.81 & 13.17 &  12 &    50.82 $\pm$ 1.01 & 1.40 $\pm$ 0.21 &  52.22 $\pm$ 1.03 &  535.2 $\pm$ 0.3 & 229.1 $\pm$ 0.06 & 48.6  &  2.50 &     & a \\
\enddata

\tablecomments{N$_\mathrm{sea}$ indicates the number of seasons
  observed, where 3-6 months of observations count as one season, for
  seasons having more than three images taken. The letter ``c''
  indicates a continuous set of observations during which multiple
  nights of data were taken in each season, whereas an ``s'' indicates
  scattered observations when one or more seasons have only a single
  night of observations. Generally, ``c'' observations are better. (a)
  Target has one or more parallaxes previously published by
  RECONS. The values here supersede those earlier
  values. Perc. Cmpn. indicates the percentage of eliminated brown
  dwarf companions to the extrasolar planet hosts, and the percentage
  of eliminated planetary-mass companions to the best case targets.}

\label{tab1}
\end{deluxetable}
\end{landscape}
\addtolength{\voffset}{-2in}
\newpage

\begin{deluxetable}{lrrrcrrrcccrrcc}
\centering
\setlength{\tabcolsep}{0.03in}
\tablewidth{0pt}
\tabletypesize{\tiny}
\tablecaption{Photometric and Spectroscopic Results}

\tablehead{
\colhead{Name}              &     
\colhead{$V$}               &
\colhead{$R$}               &
\colhead{$I$}               &
\colhead{Nights}            &
\colhead{$J$}               &
\colhead{$H$}               &
\colhead{$K\mathrm{_{S}}$}  &
\colhead{Spectral}          &
\colhead{Ref.}              &
\colhead{Mass}              &
\colhead{Trig. Dist.}       &
\colhead{Phot. Dist.}       &
\colhead{No. of}            &
\colhead{Notes}             \\
\colhead{}&     
\colhead{(mag)}&
\colhead{(mag)}&
\colhead{(mag)}&
\colhead{}&
\colhead{(mag)}&
\colhead{(mag)}&
\colhead{(mag)}&
\colhead{Type}&
\colhead{}&
\colhead{(M$\mathrm{_{\Sun}}$)}&
\colhead{(pc)}&
\colhead{(pc)}&
\colhead{Relations}&
\colhead{}\\
\colhead{(1)}                 &
\colhead{(2)}                 &
\colhead{(3)}                 &
\colhead{(4)}                 &
\colhead{(5)}                 &
\colhead{(6)}                 &
\colhead{(7)}                 &
\colhead{(8)}                 &
\colhead{(9)}                 &
\colhead{(10)}                &
\colhead{(11)}                &
\colhead{(12)}                &
\colhead{(13)}                &
\colhead{(14)}                &
\colhead{(15)}                }

\startdata
\multicolumn{15}{c}{Extrasolar Planet Hosts}\\
\hline
GJ 317            & 12.01 & 10.84 &  9.37 & 3 &  7.934 &  7.321 &  7.028 & M3.5 V  & 1 & 0.35 & 15.26 $\pm$  0.36 &  9.70 $\pm$ 1.53 & 12 &   \\   
BD $-$10  3166      & 10.03 &  9.58 &  9.19 & 3 &  8.611 &  8.300 &  8.124 & K3.0 V  & 2 & 0.85 & 65.19 $\pm$ 13.64 &                  &    & a \\  
GJ 581            & 10.56 &  9.44 &  8.03 & 3 &  6.706 &  6.095 &  5.837 & M3.0 V  & 2 & 0.30 &  6.30 $\pm$  0.06 &  6.60 $\pm$ 1.03 & 12 &   \\  
GJ 1214           & 14.71 & 13.27 & 11.50 & 3 &  9.750 &  9.094 &  8.782 & M4.5 V  & 1 & 0.14 & 14.27 $\pm$  0.24 & 12.42 $\pm$ 2.00 & 12 &   \\  
GJ 667C          & 10.34 &  9.29 &  8.09 & 3 &  6.848 &  6.322 &  6.036 & M1.5 V  & 2 & 0.36 &  7.10 $\pm$  0.10 &  9.41 $\pm$ 1.49 & 12 &   \\  
GJ 849            & 10.38 &  9.27 &  7.87 & 3 &  6.510 &  5.899 &  5.594 & M3.0 V  & 2 & 0.42 &  8.62 $\pm$  0.15 &  5.73 $\pm$ 0.92 & 12 &   \\  
GJ 876            & 10.18 &  8.97 &  7.40 & 3 &  5.934 &  5.349 &  5.010 & M3.5 V  & 2 & 0.27 &  4.69 $\pm$  0.09 &  3.46 $\pm$ 0.54 & 12 &   \\    
\hline																									     
\multicolumn{15}{c}{Best Case Targets}\\																				     
\hline																									     
GJ 1061           & 13.09 & 11.45 &  9.47 & 6 &  7.523 &  7.015 &  6.610 & M5.0 V  & 3 & 0.11 &  3.69 $\pm$  0.02 &  3.55 $\pm$ 0.60 & 12 &   \\
LP 944-020        & 18.69 & 16.39 & 13.98 & 3 & 10.725 & 10.017 &  9.548 & M9.0 V  & 4 & 0.08 &  6.42 $\pm$  0.04 &  7.04 $\pm$ 1.32 & 11 &   \\
GJ 1128           & 12.74 & 11.36 &  9.62 & 3 &  7.953 &  7.385 &  7.037 & M4.0 V  & 2 & 0.15 &  6.48 $\pm$  0.03 &  6.33 $\pm$ 1.00 & 12 &   \\
DENIS J1048-3956  & 17.37 & 14.98 & 12.47 & 4 &  9.538 &  8.905 &  8.447 & M8.0 V  & 2 & 0.08 &  4.03 $\pm$  0.01 &  4.48 $\pm$ 0.73 & 10 &   \\
SCR 1138-7721     & 14.78 & 13.20 & 11.24 & 4 &  9.399 &  8.890 &  8.521 & M5.0 V  & 3 & 0.11 &  8.31 $\pm$  0.07 &  9.45 $\pm$ 1.71 & 12 &   \\
Proxima Cen       & 11.13 &  9.45 &  7.41 & 3 &  5.357 &  4.835 &  4.384 & M5.0 V  & 2 & 0.11 &  1.30 $\pm$  0.01 &  1.15 $\pm$ 0.18 & 12 & b \\
\hline																									    
\multicolumn{15}{c}{Confirmed Binaries}\\																				    
\hline																								    
LHS 1582AB       & 14.69J & 13.33J & 11.60J & 4 &  9.799J &  9.177J &  8.854J & M4.5 VJ & 1&      & 19.67 $\pm$  0.47 & 13.27 $\pm$ 2.25 & 12 & c \\
GJ 748AB         & 11.10J &  9.95J &  8.47J & 3 &  7.087J &  6.572J &  6.294J & M3.5 VJ & 2&      & 10.00 $\pm$  0.12 &  7.69 $\pm$ 1.26 & 12 & c \\
LHS 3738AB       & 15.78J & 14.29J & 12.46J & 3 & 10.654J & 10.091J &  9.761J & M4.5 VJ & 5&      & 19.15 $\pm$  0.38 & 18.50 $\pm$ 2.96 & 12 & c \\
\enddata

\tablecomments{(a) Distance estimate only applicable to M dwarfs; (b)
  Actual error is $\pm$ 0.002 pc in trigonometric distance; (c) ``J''
  signifies joint photometry and spectroscopy for unresolved
  binaries.}  \tablerefs{For spectral types: (1) \citet{rei95}; (2)
  This work; (3) \citet{hen06}; (4) \citet{die13}; (5) \citet{haw96}}
\label{tab2}

\end{deluxetable}

Nine of the 16 targets in this paper have parallaxes previously
published by RECONS, and are noted in Column 18. The results presented
here supersede those published previously by RECONS because additional
data and improved reduction techniques have been used, as discussed in
detail in \citet{sub09}. The identical parallax of LP 944-020 is also
presented in \citet{die13} as part of a study of the stellar hydrogen
burning limit. For BD $-$10 3166, we did not run simulations because
it is too massive and far away for us to detect any type of substellar
companion. However, we do provide the first accurate parallax, and
conclude that BD $-$10 3166 is not physically related to the star with
a similar proper motion, LP 731-076, that is $20\arcsec$ away
\citep{barXX}.

Photometric and spectroscopic results are provided in Table
\ref{tab2}. $VRI$ photometry was taken using
the CTIO 0.9m (number of nights of photometry in Column 5), with
errors in $VRI \lesssim 0.03$ mag \citep{win11}. $JHK$ photometry was
retrieved from the Two Micron All Sky Survey (2MASS, \citet{skr06})
catalog. Spectral types are given in Column 9 with references in
Column 10. Mass estimates were calculated as discussed in
$\S$\ref{sec:analysis}. The photometric distances are calculated using
the $VRIJHK$ distance relations (number of relations in Column 14)
detailed in \citet{hen04}. For systems with photometric and
trigonometric distances that agree within the errors, we conclude that
they lack nearly equal luminosity companions. The trigonometric
distances of GJ 748AB and LHS 1582AB are greater than their
photometric distances due to companion contributions to the systems'
total flux. The two distances of LHS 3738AB agree well, indicating
that the companion is significantly fainter than the primary. GJ 317
and GJ 849 have discrepant (at 3.5$\sigma$ and 3.1$\sigma$,
respectively) photometric and trigonometric distances, which does not
necessarily mean that these stars have stellar companions, as main
sequence stars within the same spectral type can vary somewhat in
luminosity.

\citet{bon13a} note a radial velocity drift in their observations of GJ
849. \citet{mon13} also note this drift, and constrain the minimum
companion mass to $M\sin{i} < 2.5 \, \mathrm{M_{Jup}}$. Our astrometry
would show a photocenter shift for unequal mass components, as
discussed in $\S$\ref{sec:observe}. Only components of roughly equal
luminosity and mass would provide the additional flux with no
perturbation. Such a companion would have been observed to separations
as close as 1{\arcsec} in our images, which corresponds to $\sim$9
AU. At a semimajor axis of 9 AU, the orbital period is
29.5 years for an equal mass companion. This results in velocities
for each component 9.1 km/s for edge-orbits. Thus, for most orbital
inclinations, such a stellar companion is ruled out by the radial
velocity data. Therefore, it is unlikely that a stellar companion
similar to the primary is contributing to the overluminosity we
observe.

\begin{figure*}[ht!]
\centering

\includegraphics[width=5.25in]{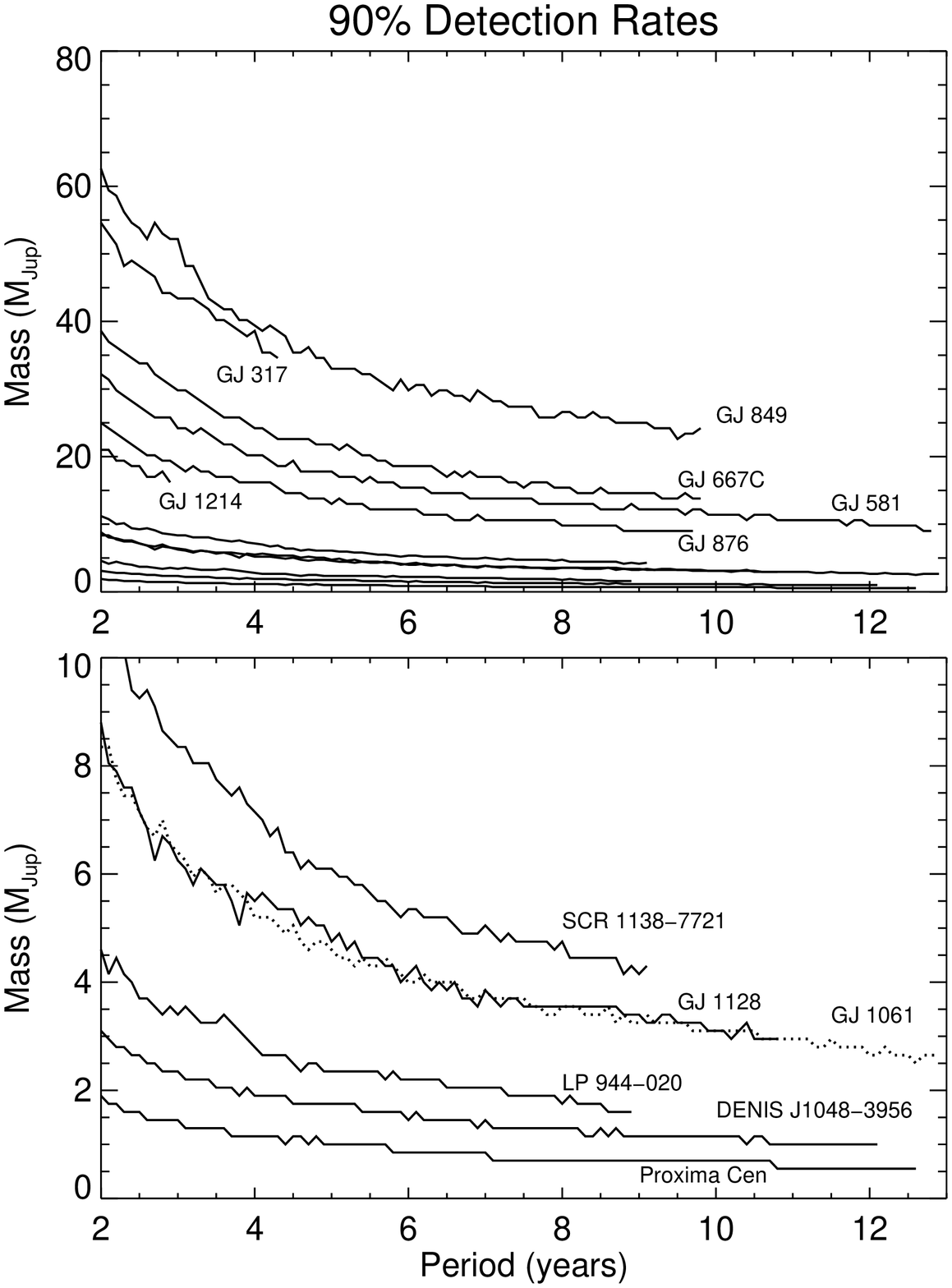}

\caption{The masses and periods at which 90\% of companions would have
  been detected, based on simulations of 10 million companions for
  each star. Companions above the lines would have been detected. The
  lines end at the length of observations for each target. The six M
  dwarfs with known planets are labeled in the top panel. The bottom
  panel is a zoom of the top, showing in greater detail the six
  additional M dwarfs that are more favorable to the astrometric
  detection of planets. GJ 1061 is shown as a dotted line to
  differentiate it from GJ 1128. Note the different vertical scales,
  where the top panel represents primarily brown dwarfs and the bottom
  panel Jovian planets.}
\label{fig4}
\end{figure*}

Figure \ref{fig4} shows the range of periods and masses for
which 90\% of simulated companions would have been detected, based on
the simulations for objects with masses from 0.5 to 80
$\mathrm{M_{Jup}}$. As discussed in $\S$\ref{sec:analysis}, the masses,
periods, and orbital parameters of the simulated companions were
assigned randomly from a uniform distribution. For this discussion we
set the dividing line between planets and brown dwarfs at 13
$\mathrm{M_{Jup}}$, and the dividing line between brown dwarfs and
stars at 80 $\mathrm{M_{Jup}}$. The bottom panel of Figure
\ref{fig4} is an inset of the top, showing the best case targets
in more detail. The noisy nature of the lines is due to the sizes of
the bins used in the simulations. The bin sizes were chosen to achieve
a reasonably high resolution, while still having enough simulated
companions in each bin. The minimum detectable companion mass is
smallest for companions with long periods, which produce the largest
amplitude perturbations in the astrometric data. The lower panel
indicates that for the best case targets (stars at close distances and
of low mass). We are most sensitive to Jovian-type planets in
Jovian-type orbits.
  
For a companion at a given mass and orbital period, the amplitude of
the resulting astrometric perturbation depends on the orbital
parameters of the system, and the mass and distance of the
primary. The detection limits we report are based on simulations of
companions with a wide range of masses, periods, and orbital
parameters. Therefore we give a few representative examples of how the
results in Figure \ref{fig4} translate into astrometric perturbations
in mas. In the case of a face-on, circular orbit, a 20
$\mathrm{M_{Jup}}$ companion in a 4 year orbit around GJ 581 would
cause a 16 mas perturbation, while a 15 $\mathrm{M_{Jup}}$ companion
in an 8 year orbit would cause a 20 mas perturbation, and a 10
$\mathrm{M_{Jup}}$ companion in a 12 yr orbit would cause a 17 mas
perturbation. For circular, face-on orbits around Proxima Centauri,
companions of 1.5, 1, and 0.5 $\mathrm{M_{Jup}}$ in orbits of 4, 8,
and 12 years would cause perturbations of 12, 13, and 8 mas,
respectively. These values are significantly greater than the per
observation precisions listed Table \ref{tab1} --- 7.93 mas for GJ 581
and 4.83 mas for Proxima. Thus, the 90\% detection thresholds given in
Figure \ref{fig4} are reasonable

For the four planet hosts observed longer than eight years, Column 17
of Table \ref{tab1} gives the percentages of simulated
\textit{brown dwarf} companions, ranging from 81 -- 99\%, eliminated
with orbits between two and eight years, and masses between 13 and 80
$\mathrm{M_{Jup}}$. Approximately 92\% of all simulated brown dwarfs
have been eliminated as companions to those stars known to host
exoplanets. For the six more astrometrically favorable targets, we
calculate the percentages of simulated \textit{planetary} companions
eliminated with orbits between two and eight years and masses between
1 and 13 $\mathrm{M_{Jup}}$, with results ranging from 69 -- 99\%.  We
have eliminated $\sim$86\% of all simulated planets with masses of 1
-- 13 $\mathrm{M_{Jup}}$ around these six astrometrically favorable
stars, and effectively all brown dwarf companions in orbital periods
of 2 -- 8 years.

\begin{figure*}[ht!]
\centering

{\includegraphics[scale=0.28]{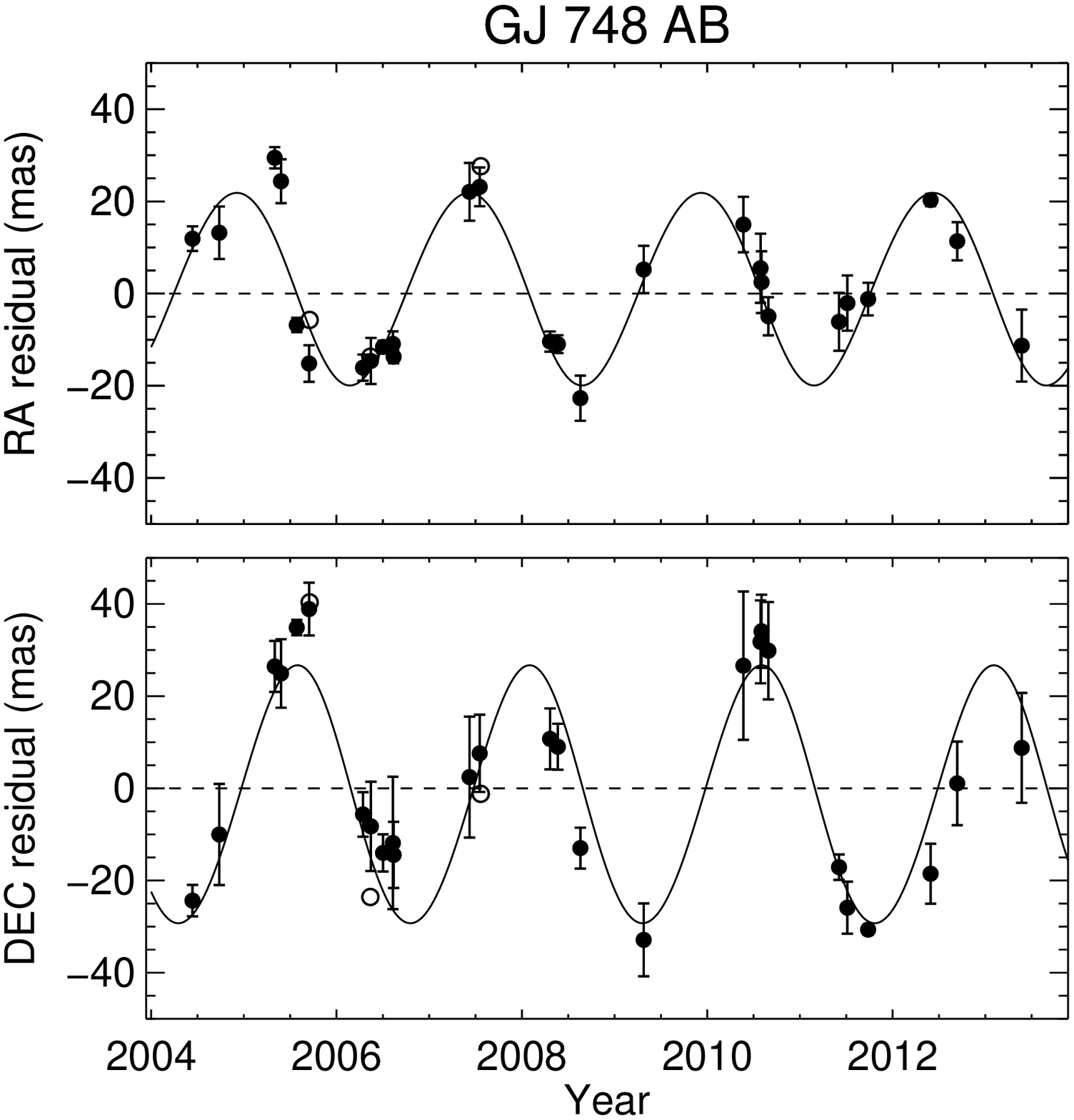}} 
\subfigure
{\includegraphics[scale=0.28]{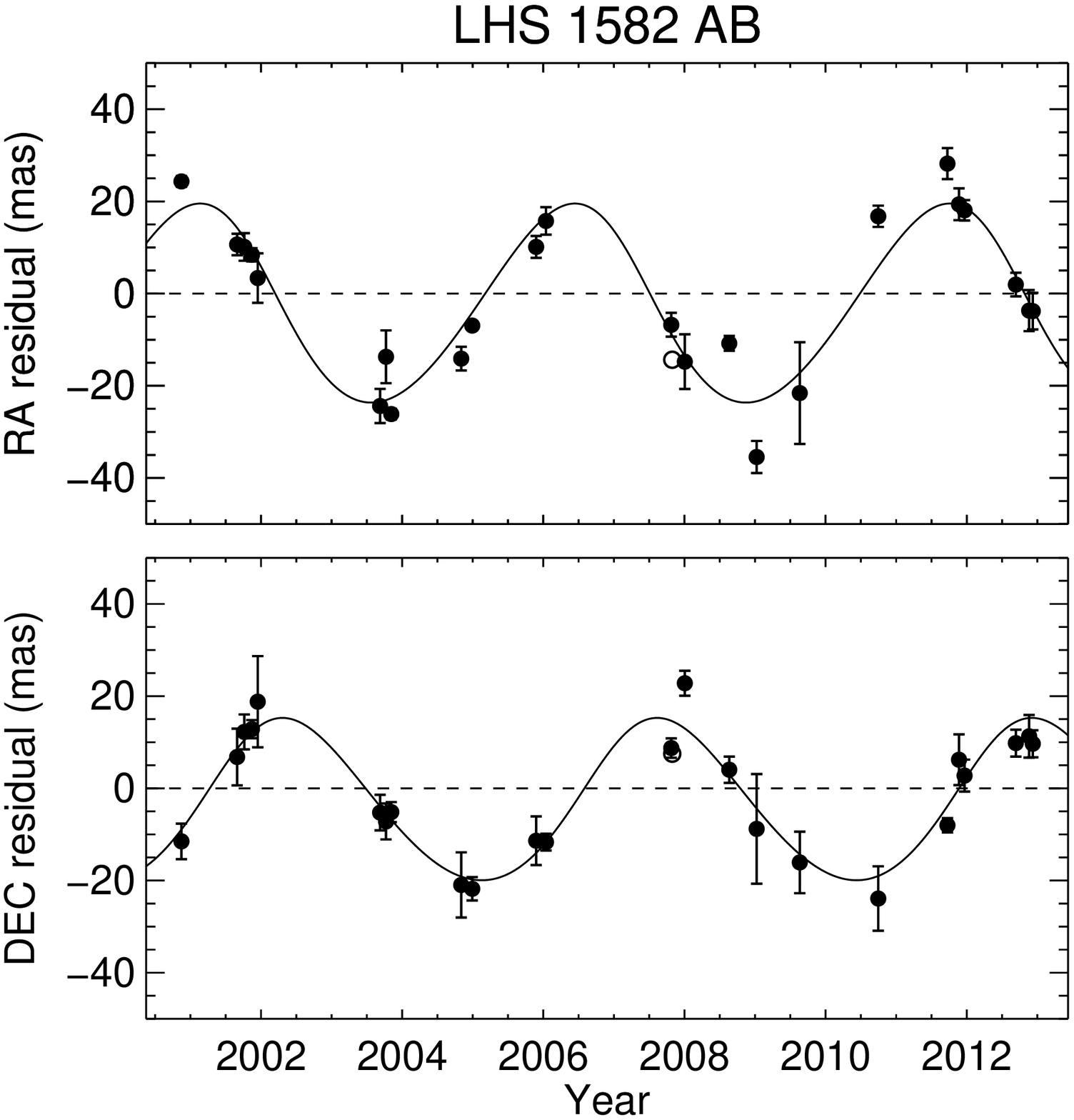}}
\subfigure
{\includegraphics[scale=0.28]{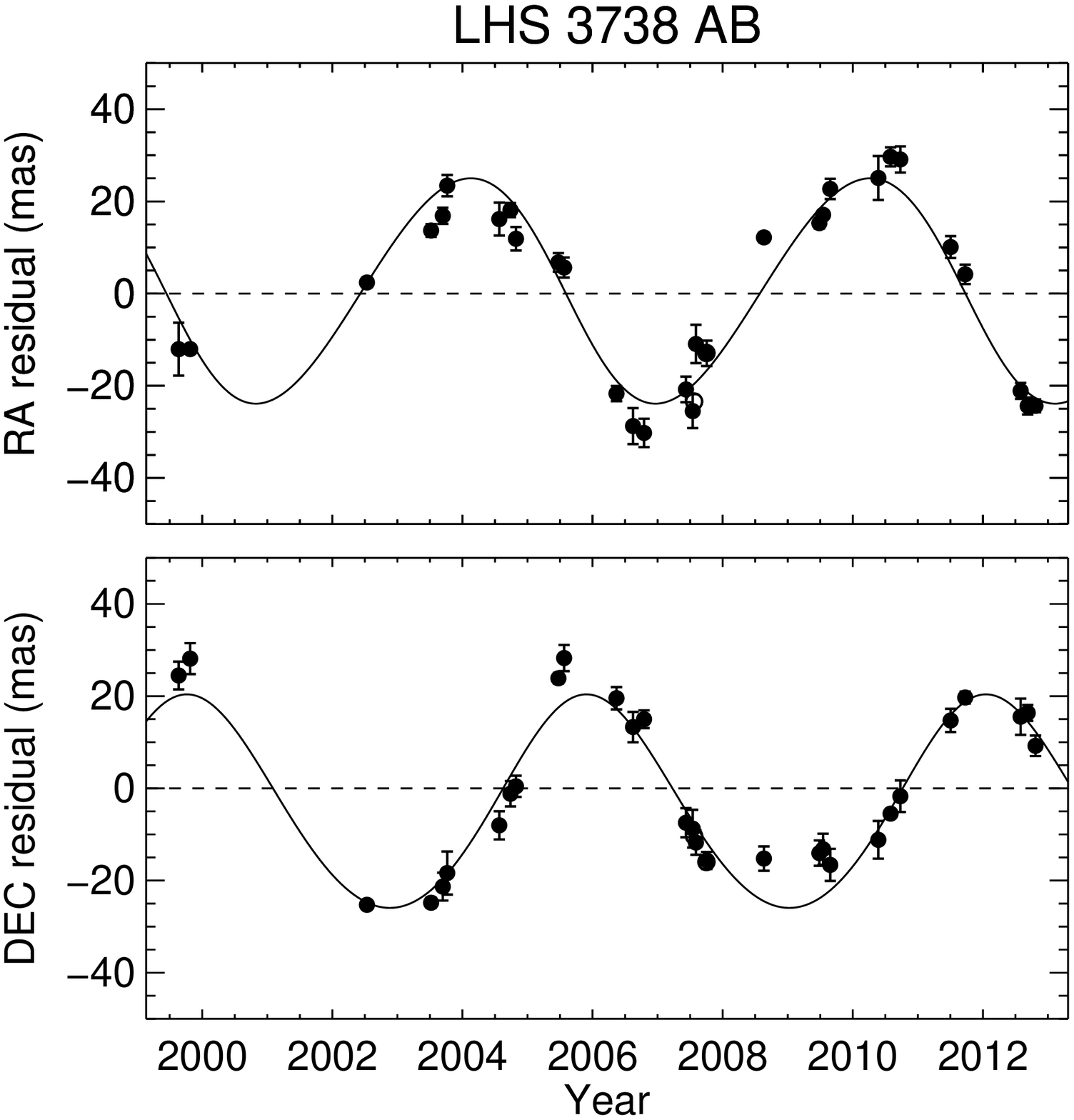}}

\vspace{0.1in}
{\includegraphics[scale=0.28]{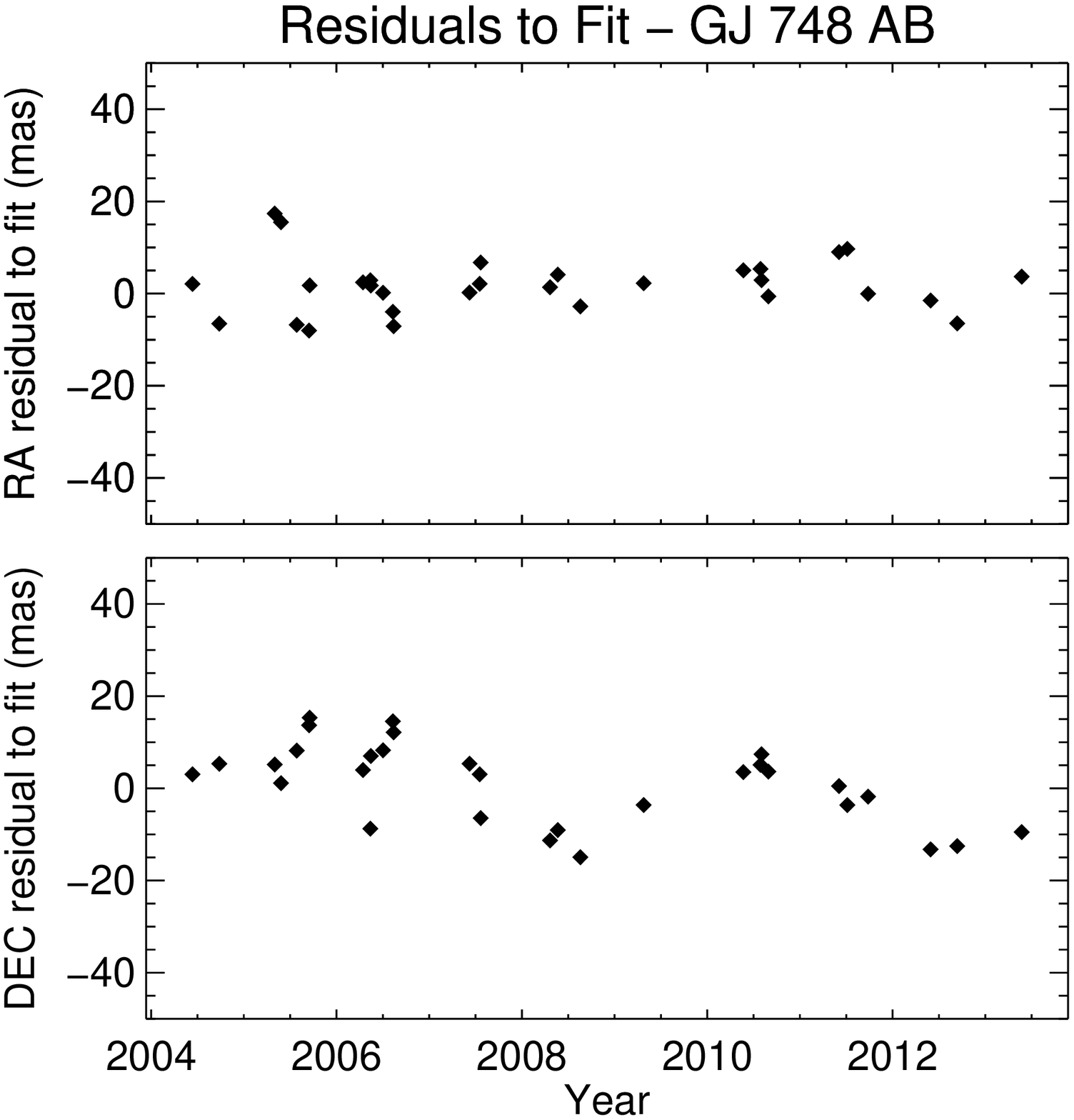}}
\subfigure
{\includegraphics[scale=0.28]{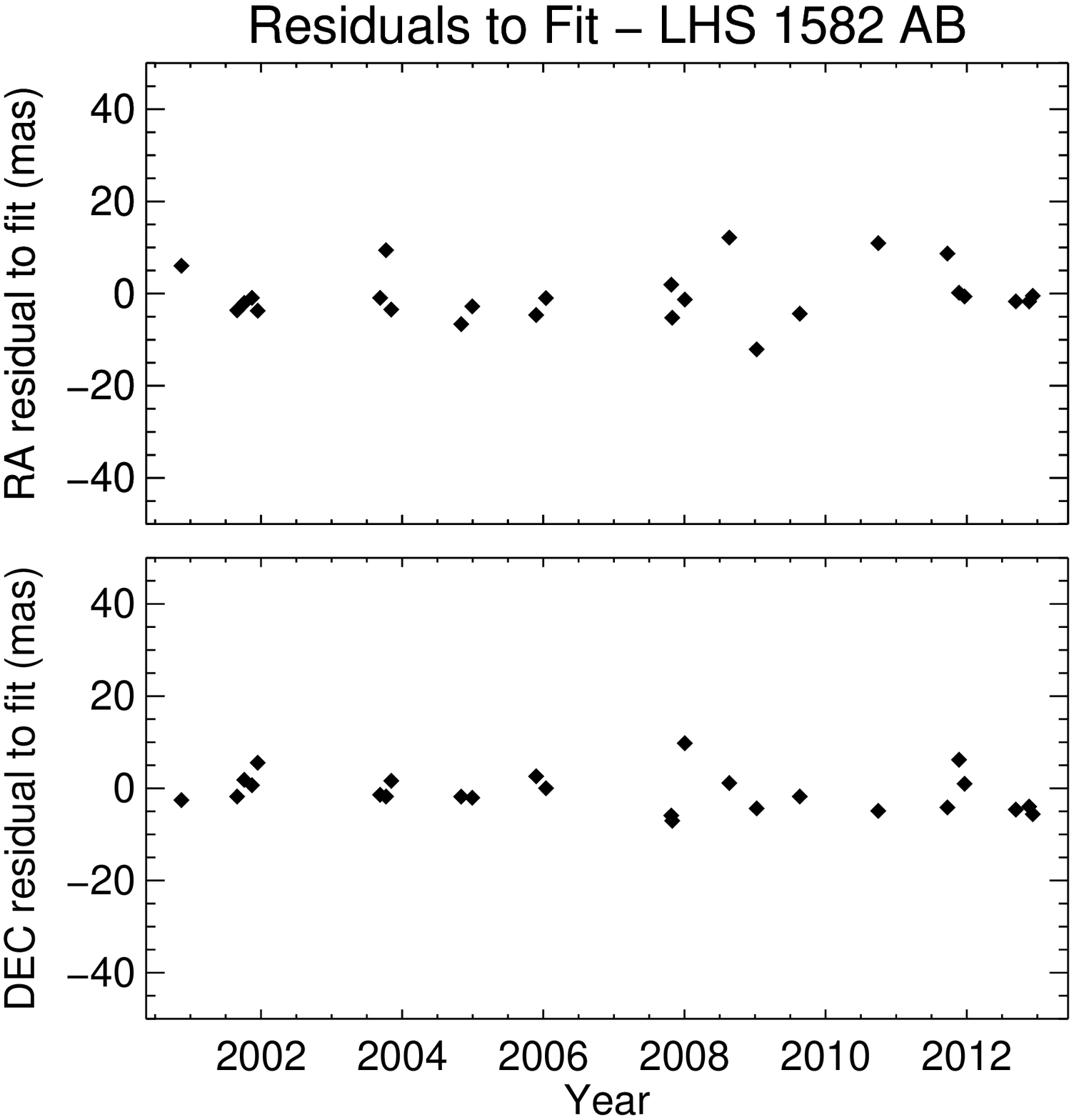}}
\subfigure
{\includegraphics[scale=0.28]{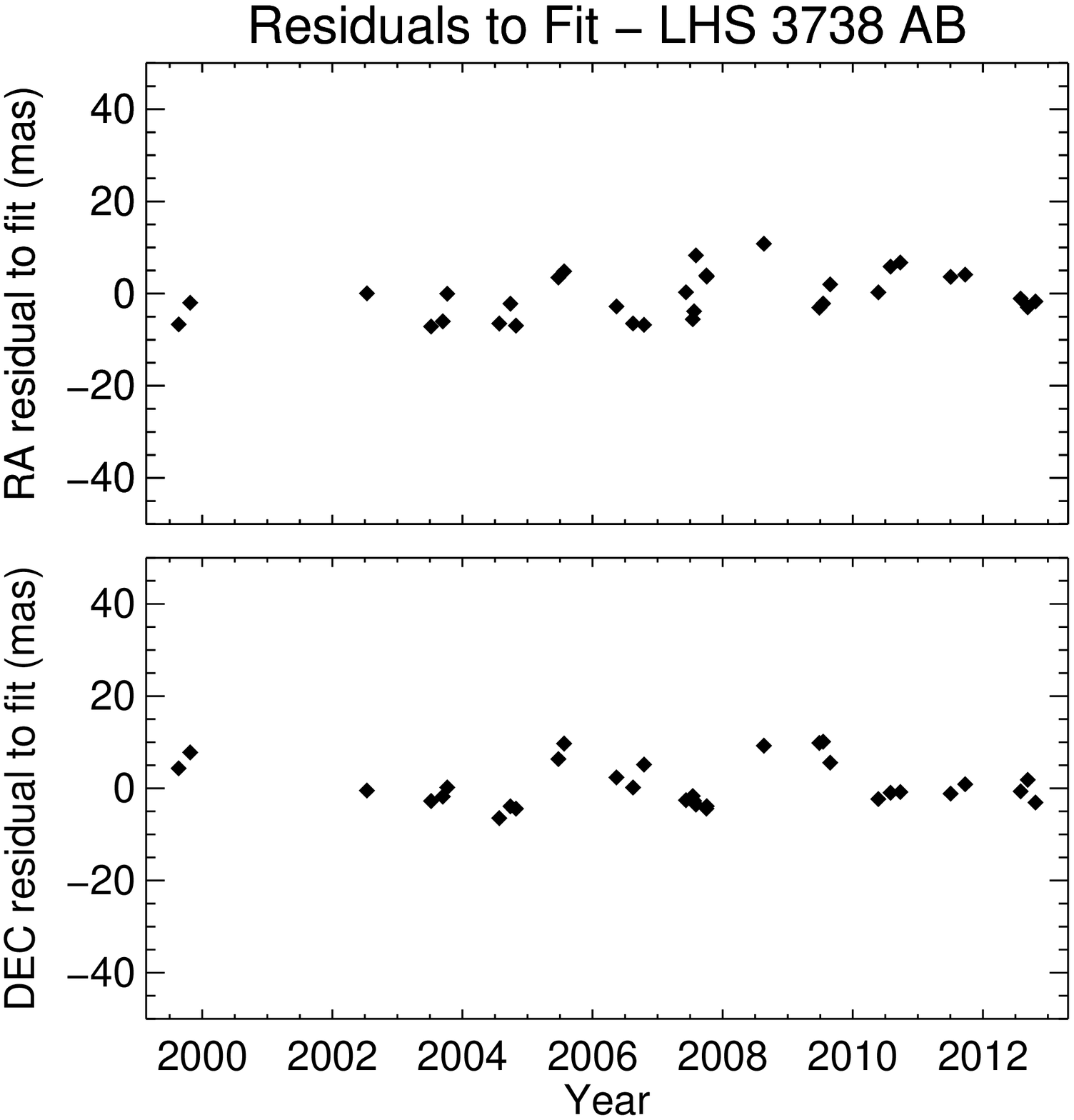}}

\caption{The top panels show the orbital fits for the astrometric
  residuals of three confirmed binaries. Filled circles represent the
  mean of typically three to five frames taken in a single night. Open
  circles represent nights for which there is only one frame, and are
  not used to derive the fit. The bottom panels show the residuals to
  the orbital fits.  All panels are on a $\pm$50 mas scale.}
\label{fig5}

\end{figure*}

Photocentric orbital solutions for the three binaries are shown in
Figure \ref{fig5} with the corresponding orbital parameters
given in Table \ref{tab3}. From our astrometric data for GJ 748
AB, we find an orbital period of 2.504 $\pm$ 0.025 years, which is
consistent with the two detailed studies of the system by
\citet{fra99}, who found P = 2.466 $\pm$ 0.008 years, and
\citet{ben01}, who found P = 2.469 $\pm$ 0.001 years using HST Fine
Guidance Sensor data.  However, we determine an eccentricity of 0.06,
which is inconsistent with the value of 0.45 found in both of the HST
studies. We utilized the orbit-fitting code described in \citet{hrt89}
and set starting eccentricities of 0.05 to 0.95 in increments of 0.05;
regardless, our data converged to the e = 0.06 value each time.

\begin{deluxetable}{lrrrrrrr}
\centering
\tablewidth{0pt}
\setlength{\tabcolsep}{0.06in}
\tabletypesize{\small}
\tablecaption{Orbital Parameters of Known Binaries}

\tablehead{
\colhead{Name}&     
\colhead{P}&
\colhead{T$_{0}$}&
\colhead{a$_{\mathrm{phot}}$}&
\colhead{e}&
\colhead{i}&
\colhead{Long. Peri. ($\omega$)}& 
\colhead{Long. Node ($\Omega$)}\\
\colhead{}&
\colhead{(years)}&
\colhead{}&
\colhead{(mas)}&
\colhead{}&
\colhead{(deg)}&
\colhead{(deg)}&
\colhead{(deg)}\\
\colhead{(1)}                 &
\colhead{(2)}                 &
\colhead{(3)}                 &
\colhead{(4)}                 &
\colhead{(5)}                 &
\colhead{(6)}                 &
\colhead{(7)}                 &
\colhead{(8)}                 }

\startdata
GJ 748AB   & 2.504$\pm$0.025 & 2005.86$\pm$0.25 & 28.1$\pm$1.6 & 0.06$\pm$0.04 & 137.8$\pm$9.0 & 218.2$\pm$40.3 & 173.6$\pm$12.1 \\

LHS 1582AB & 5.309$\pm$0.049 & 2001.84$\pm$0.14 & 21.9$\pm$1.3 & 0.17$\pm$0.03 & 143.6$\pm$7.5 &  62.0$\pm$14.3 &  97.9$\pm$12.4 \\

LHS 3738AB & 6.141$\pm$0.059 & 2005.73$\pm$0.16 & 28.1$\pm$1.2 & 0.12$\pm$0.02 & 131.8$\pm$4.1 & 130.5$\pm$11.2 & 130.5$\pm$ 5.1 \\

\enddata

\label{tab3}
\end{deluxetable}

The discrepancy between our eccentricity and that of the HST studies
is likely due to our observations of GJ 748 AB having been taken at
the two different $V$ filters discussed in $\S$ 2.1. While the two
filters are photometrically identical within measurable errors
\citep{jao11}, they are not astrometrically identical. We have
analyzed the astrometric residuals for over 500 targets without
detectable perturbations in the three different filters ($V,R,$ and
$I$) over the length of our observing program. The R and I filters are
stable, but astrometric offsets in the V filters are evident over the
time period when the problematic ``new'' $V$ filter was used. These
offsets have been mitigated as discussed in $\S$ 2.1, allowing data
from both $V$ filters to be used to produce reliable parallax
results. In the case of GJ 748 AB, we are able to recover the correct
period, but the offsets in the residuals are likely contributing to
the errant eccentricity. We presently do not possess enough
observations to perform a reduction of GJ 748 AB without the
problematic $V$ filter data. We include the current solution because
it is our only system for which an accurate period has been published,
against which to compare our results.

In contrast to our photocenter data, the FGS observations resolve the
system into two components at 15 epochs over 1.8 years. They are
exquisitely sensitive to the separation and position angle of the
secondary from the primary, and are to be preferred to our
ground-based results for the eccentricity. A clever suggestion by Hugh
Harris of USNO has been suggested to solve this dilemma. Because in an
unresolved system the center of mass location is unknown, the zero
points for the residuals in R.A. and Decl. are unknown. By shifting
the zero points and fitting the residuals, a different eccentricity
may be derived. We await the acquisition of resolved data for several
more systems before exploring this technique so that a robust analysis
can be accomplished. At present errors on the eccentricities in Table
\ref{tab3} should be treated with caution. The orbital solutions for
LHS 1582AB and LHS 3738AB are updated and improved over those
presented in \citet{rie10}, which were the first orbits presented for
each system. In addition to demonstrating the astrometric detection
and characterization of unresolved companions, these results can
provide additional dynamically determined masses for M dwarfs, once
the systems have been resolved.


\section{Discussion}
\label{sec:discuss}
In narrow-angle field astrometry programs such as the one outlined
here, the lowest mass companions detectable by astrometry are
dependent on a number of factors. These include the apparent
brightness of the host star and availability of suitable reference
stars, which affect the precision of the astrometric
measurements. Additionally, the target star mass, companion mass, the
luminosity ratio, orbital separation, and system distance determine
the size of the astrometric perturbation. The M dwarfs with planets
that have been discovered so far, including the six in this paper, are
not the most favorable to our astrometric observations. They are among
the brighter red dwarfs, with the latest type stars at M3.5V for the
radial velocity detections included in this paper. We are rarely
sensitive to planets around these stars. However, we are able to rule
out the presence of a large fraction of potential brown dwarf
companions with masses of $13 - 80$ $\mathrm{M_{Jup}}$. As
observations continue, we will be able to rule out companions at
longer periods, particularly in the cases of GJ 317 and GJ 1214, which
we have so far only observed for 3 -- 4 years. In addition, we have
demonstrated that we are sensitive to planets with masses of $1 - 13$
$\mathrm{M_{Jup}}$ around M dwarfs that are more favorable to
astrometric observations. Of particular interest, we find that Proxima
Centauri has no companions more massive than 2 $\mathrm{M_{Jup}}$ with
periods of 2 -- 5 years and more massive than 1 $\mathrm{M_{Jup}}$ for
5 -- 12 years. This is to be contrasted to the results of other
companion searches of Proxima Centauri. Using HST Fine Guidance Sensor
data, \citet{ben99} report a companion detection sensitivity of 1
$\mathrm{M_{Jup}}$ at a 60 day period. Based on 7 years of RV
observations, \citet{end08} preclude the presence of companions with
$M\sin(i) \ge 1\,\mathrm{M_{Nep}}\,(0.05\,\mathrm{M_{Jup}})$ at
periods $\le 2.7$ years. Together these studies eliminate all Jupiter
mass planets around Proxima Centauri for orbital periods out to 12
years.

\begin{deluxetable}{lccrcccrccc}
\centering
\setlength{\tabcolsep}{0.05in}
\tabletypesize{\footnotesize}
\tablewidth{0pt}
\tablecaption{Planets Orbiting M Dwarfs Within 25 pc}

\tablehead{
\colhead{Name}&     
\colhead{R.A. (J2000.0)}&
\colhead{Decl. (J2000.0)}&
\colhead{$\pi_{\mathrm{mean}}$}&
\colhead{N$_\pi$}&
\colhead{Ref.}&
\colhead{M$\sin(i)$}&
\colhead{P}&
\colhead{Ref.}&
\colhead{Notes}\\
\colhead{}&
\colhead{}&
\colhead{}&
\colhead{(mas)}&
\colhead{}&
\colhead{}&
\colhead{(M$_{\mathrm{Jup}}$)}&
\colhead{(years)}&
\colhead{}&
\colhead{}\\
\colhead{(1)}                 &
\colhead{(2)}                 &
\colhead{(3)}                 &
\colhead{(4)}                 &
\colhead{(5)}                 &
\colhead{(6)}                 &
\colhead{(7)}                 &
\colhead{(8)}                 &
\colhead{(9)}                 &
\colhead{(10)}                 }

\startdata

GJ 163 b     & 04 09 15.6 & $-$53 22 25 &  66.59 $\pm$ 1.79 & 2 & 1,2     & 0.03  & 0.024 & 9  &   \\	     
GJ 163 c     &            &             &                   &   &         & 0.02  & 0.070 & 9  &   \\	     
GJ 163 d     &            &             &                   &   &         & 0.09  & 1.654 & 9  &   \\	     
GJ 176 b     & 04 42 55.7 & $+$18 57 29 & 110.00 $\pm$ 2.00 & 2 & 1,2     & 0.03  & 0.024 & 10  &   \\	     
GJ 179 b     & 04 52 05.7 & $+$06 28 35 &  80.82 $\pm$ 3.78 & 2 & 1,2     & 0.82  & 6.264 & 11 &   \\
GJ 317 b     & 08 40 55.7 & $-$23 28 00 &  65.34 $\pm$ 0.39 & 3 & 1,3,4   & 1.81  & 1.895 & 4  & a \\ 
LHS 2335 b   & 10 58 35.0 & $-$31 08 38 &  50.55 $\pm$ 1.55 & 1 & 5       & 0.02  & 0.007 & 12 &   \\ 
GJ 433 b     & 11 35 26.9 & $-$32 32 23 & 112.09 $\pm$ 1.43 & 2 & 1,2     & 0.02  & 0.020 & 13 & a \\
GJ 1148 b    & 11 41 44.6 & $+$42 45 07 &  88.81 $\pm$ 2.14 & 2 & 1,2     & 0.30  & 0.113 & 14 &   \\
GJ 436 b     & 11 42 11.0 & $+$26 42 23 &  98.95 $\pm$ 2.07 & 2 & 1,2     & 0.07  & 0.007 & 15 &   \\
GJ 581 b     & 15 19 26.0 & $-$07 43 20 & 159.28 $\pm$ 1.32 & 3 & 1,2,3   & 0.05  & 0.015 & 16 &   \\
GJ 581 c     &            &             &                   &   &         & 0.02  & 0.035 & 16 &   \\
GJ 581 d     &            &             &                   &   &         & 0.02  & 0.183 & 16 & b \\
GJ 581 e     &            &             &                   &   &         & 0.01  & 0.009 & 16 &   \\
LP 804-027 b & 16 12 41.7 & $-$18 52 31 &  69.46 $\pm$ 3.12 & 1 & 2       & 2.10  & 0.306 & 17 &   \\
GJ 649 b     & 16 58 08.8 & $+$25 44 39 &  97.28 $\pm$ 1.32 & 2 & 1,2     & 0.33  & 1.638 & 18 & a \\
GJ 1214 b    & 17 15 18.9 & $+$04 57 50 &  69.04 $\pm$ 0.54 & 3 & 1,3,6   & 0.02  & 0.004 & 19 & c \\
GJ 667C  b   & 17 18 57.1 & $-$34 59 23 & 138.24 $\pm$ 0.57 & 3 & 1,3,7   & 0.02  & 0.020 & 13 & a \\
GJ 667C  c   &            &             &                   &   &         & 0.01  & 0.077 & 13 &   \\
GJ 674 b     & 17 28 39.9 & $-$46 53 42 & 220.11 $\pm$ 1.39 & 2 & 1,2     & 0.03  & 0.013 & 20 &   \\
GJ 832 b     & 21 33 33.9 & $-$49 00 32 & 202.03 $\pm$ 1.00 & 2 & 1,2     & 0.64  & 9.353 & 21 &   \\
GJ 849 b     & 22 09 40.3 & $-$04 38 26 & 113.46 $\pm$ 1.32 & 3 & 1,2,3   & 0.90  & 5.241 & 22 & a \\
GJ 876 b     & 22 53 16.7 & $-$14 15 49 & 214.45 $\pm$ 0.57 & 4 & 1,2,3,8 & 1.95  & 0.167 & 23 &   \\
GJ 876 c     &            &             &                   &   &         & 0.61  & 0.082 & 23 &   \\
GJ 876 d     &            &             &                   &   &         & 0.02  & 0.005 & 23 &   \\
GJ 876 e     &            &             &                   &   &         & 0.04  & 0.342 & 23 &   \\

\enddata

\tablecomments{To provide a uniform format, minimum masses and orbital
  periods are in some cases rounded to fewer significant digits than
  in the original publications. Recent discoveries GJ 191 b/c \citep{ang14}
  and GJ 687 b \citep{bur14} were not listed in exoplanets.org as of 2014 July
  1 (a) Additional companion(s) not listed on both exoplanet.eu and
  exoplanets.org; (b) Signal attributed to stellar activity \citep{rob14} (c)
  transiting planet; all others are radial velocity detections.}

\tablerefs{
(1) Yale Parallax Catalog, \citet{van95}; 
(2) \textit{Hipparcos}, \citet{vLe07};
(3) This work; 
(4) \citet{ang12};
(5) \citet{rie10}; 
(6) \citet{ang13}:
(7) \citet{fab00}:
(8) \citet{ben02};
(9)  \citet{bon13b};
(10) \citet{for09};
(11) \citet{how10}; 
(12) \citet{bon11};  
(13) \citet{del13}; 
(14) \citet{hag10};
(15) \citet{man07};
(16) \citet{may09};
(17) \citet{app10}; 
(18) \citet{joh10}; 
(19) \citet{har13};
(20) \citet{bon07};
(21) \citet{bai09};
(22) \citet{mon13}; 
(23) \citet{riv10}.
}
\label{tab4}

\end{deluxetable}

In the broader context, these results are consistent with recently
published searches for Jovian companions to M dwarfs at shorter
orbital periods. Transit searches are unlikely to detect companions at
Jovian orbits, due to the narrow range of detectable
inclinations. \citet{bta13} found no Jupiter-sized planets in their
transit search, which is most sensitive to companions at orbital
periods less than 10 days, and conclude that such planets rarely orbit
M dwarfs. Based on the first two years of their astrometric search,
\citet{sah14} find no planetary mass companions to the 20 M and L
dwarfs they observed. They determine the occurrence rate of planets
more massive than $\sim$5 $\mathrm{M_{Jup}}$ to have an upper limit of
9\%. 

The longer time coverage of the astrometric results presented in this
work overlap most closely with radial velocity results. \citet{bon13a}
report detection limits based on radial velocity measurements for 102
nearby M dwarfs, of which six are featured in this paper. Among the
102 M dwarfs searched, they confirm only two planets with orbital
periods longer than 100 days. At an orbital period of 1000 days (2.7
years), they report a detection sensitivity of $M\sin(i) \le 2 \,
M_{Jup}$ around 90\% of stars observed, and $M\sin(i) \le 30 \,
\mathrm{M_{Jup}}$ at 10000 days (27 years). \citet{mon13} report that
6.5\% $\pm$ 3.0\% of M dwarfs host a 1 to 13 $\mathrm{M_{Jup}}$ planet
at a separation less than 20 AU, based on their radial velocity survey
and high resolution imaging.

The results presented here are among the first astrometric searches
for Jovian companions at Jovian orbits, and fill in relatively
unexplored mass and period parameter space --- models of how M dwarf
planetary systems form and evolve must now explain the lack of massive
companions with long period orbits like those in our Solar
System. Looking forward, the \textit{Gaia} mission may detect up to
2600 planets within 100 pc \citep{soz13}. However, its ability to determine
accurate masses and orbits will be limited to orbital periods less
than 6 years, a fraction of the time coverage of our ground-based
astrometric observations.

Finding nearby M dwarfs with planets remains an important challenge,
as the closest planets are the brightest and most easily studied, and
M dwarfs dominate the stellar population.  As part of its mission to
characterize the solar neighborhood, RECONS is developing a database
of all objects with accurate trigonometric parallaxes placing them
within 25 pc ($\pi_{trig} \ge 40$ mas with an error $\le 10$ mas). For
an extrasolar planet to be included in the RECONS Database it must
orbit a star that meets the above criteria, or be a free-floating
object with a comparable parallax, and be listed in \textit{both} the
Extrasolar Planets Encyclopaedia (\textit{exoplanet.eu}) and the
Exoplanet Orbit Database (\textit{exoplanets.org}, \citet{wri11}). The
RECONS Database currently contains 1074 systems having M dwarf
primaries ($9.0 \le$ $M_V \le 21.0$) within 25 pc, only 17 of which
have detected exoplanets as of 2014 July 1, listed in Table
\ref{tab4}. The error-weighted mean parallax for each system is given
in Column 4, including the parallaxes in this work and published
values. Minimum masses and orbital periods for planets with references
are listed in Columns 7 -- 9. Note that only 3 of the 26 reported
planets have masses greater than Jupiter's. Based on the 36 M dwarf
primaries within 5 pc \citep{hen13}, we anticipate that there are 4500
M dwarf primaries within 25 parsecs, yet only 17 so far have been
found to host planets. Clearly, many planets lurk undetected in the
solar neighborhood. Discovering these planets will require a wide
variety of survey techniques, and as sensitivities are improved,
astrometry will continue to play an important role.


\acknowledgements 

We wish to thank Cassy Davison for her assistance developing the
simulations, and Sergio Dieterich for helpful conversations regarding
brown dwarf luminosities, and his assistance in surveying the existing
literature. We also wish to thank Hugh Harris for his helpful
suggestion regarding the eccentricities of unresolved binaries. This
effort has been supported by the National Science Foundation via
grants AST 05-07711 and AST 09-08402, and the long-term cooperative
efforts of the National Optical Astronomy Observatories and the
members of the SMARTS Consortium. This research has made use of the
Exoplanet Orbit Database and the Exoplanet Data Explorer at
\textit{exoplanets.org}, as well as the Extrasolar Planet Encyclopaedia at
\textit{exoplanet.eu}. This research has also made use of Aladin and the SIMBAD
database, operated at CDS, Strasbourg, France.

{\it Facilities: \facility{CTIO, 2MASS}}


\clearpage


\end{document}